\documentclass[10pt,a4paper,twoside]{article}

\DeclareMathAlphabet{\mathsc}{OT1}{cmr}{m}{sc}
\def\testbx{bx}%
\DeclareRobustCommand{\ion}[2]{%
\relax\ifmmode
\ifx\testbx\f@series
{\mathbf{#1\,\mathsc{#2}}}\else
{\mathrm{#1\,\mathsc{#2}}}\fi
\else\textup{#1\,{\mdseries\textsc{#2}}}%
\fi}

\def\la{\mathrel{\mathchoice {\vcenter{\offinterlineskip\halign{\hfil
$\displaystyle##$\hfil\cr<\cr\sim\cr}}}
{\vcenter{\offinterlineskip\halign{\hfil$\textstyle##$\hfil\cr
<\cr\sim\cr}}}
{\vcenter{\offinterlineskip\halign{\hfil$\scriptstyle##$\hfil\cr
<\cr\sim\cr}}}
{\vcenter{\offinterlineskip\halign{\hfil$\scriptscriptstyle##$\hfil\cr
<\cr\sim\cr}}}}}
\def\ga{\mathrel{\mathchoice {\vcenter{\offinterlineskip\halign{\hfil
$\displaystyle##$\hfil\cr>\cr\sim\cr}}}
{\vcenter{\offinterlineskip\halign{\hfil$\textstyle##$\hfil\cr
>\cr\sim\cr}}}
{\vcenter{\offinterlineskip\halign{\hfil$\scriptstyle##$\hfil\cr
>\cr\sim\cr}}}
{\vcenter{\offinterlineskip\halign{\hfil$\scriptscriptstyle##$\hfil\cr
>\cr\sim\cr}}}}}
\def\degr{\hbox{$^\circ$}}

\def\utw{\smash{\rlap{\lower5pt\hbox{$\sim$}}}}
\def\udtw{\smash{\rlap{\lower6pt\hbox{$\approx$}}}}

\def\diameter{{\ifmmode\mathchoice
{\ooalign{\hfil\hbox{$\displaystyle/$}\hfil\crcr
{\hbox{$\displaystyle\mathchar"20D$}}}}
{\ooalign{\hfil\hbox{$\textstyle/$}\hfil\crcr
{\hbox{$\textstyle\mathchar"20D$}}}}
{\ooalign{\hfil\hbox{$\scriptstyle/$}\hfil\crcr
{\hbox{$\scriptstyle\mathchar"20D$}}}}
{\ooalign{\hfil\hbox{$\scriptscriptstyle/$}\hfil\crcr
{\hbox{$\scriptscriptstyle\mathchar"20D$}}}}
\else{\ooalign{\hfil/\hfil\crcr\mathhexbox20D}}%
\fi}}

\def\aj{AJ}%
\def\araa{ARA\&A}%
\def\apj{ApJ}%
\def\apjl{ApJ}%
\def\aap{A\&A}%
\def\aaps{A\&AS}%
\usepackage{epsfig}
\usepackage{baltlat5}
\usepackage{wrapfig}
\begin{document}
   \ \
   \vspace{-0.5mm}

   \setcounter{page}{1}
   \vspace{-2mm}

   \titlehead{Baltic Astronomy, vol.\ts xx, xxx--xxx, 2006.}

   \titleb{GAUSSIAN DECOMPOSITION OF \ion{H}{i} SURVEYS -- II.\\
           SEPARATION OF PROBLEMATIC GAUSSIANS}

   \begin{authorl}
      \authorb{U.~Haud}{1}
      \authorb{P.~M.~W.~Kalberla}{2}
   \end{authorl}

   \begin{addressl}
      \addressb{1}{Tartu Observatory, EE61602 T\~oravere, Tartumaa, Estonia}
      \addressb{2}{Radioastronomisches Institut der Universit\"at Bonn, Auf
                   dem H\"ugel 71, 53121 Bonn, Germany}
   \end{addressl}

   \submitb{Received \today; revised \today}

   \begin{summary}
      We have analyzed the results of the Gaussian decomposition of the
      Leiden/Dwingeloo Survey (LDS) of galactic neutral hydrogen for
      the presence of Gaussians probably not directly related to
      galactic \ion{H}{i} emission. It is demonstrated that at least
      three classes of such components can be distinguished. The
      narrowest Gaussians, obtained during the decomposition, mostly
      represent stronger random noise peaks in profiles and some still
      uncorrected radio-interferences. Many of slightly wider weak
      Gaussians are caused by increased uncertainties near the profile
      edges and with the still increasing width the baseline problems
      become dominating among weak components. Statistical criteria are
      given for separation of the parameter space regions of the
      Gaussians, most likely populated with the problematic components
      from those where the Gaussians are with higher probability
      describing the actual Milky Way \ion{H}{i} emission.

      The same analysis is applied also to the Leiden/Argentina/Bonn
      survey (LAB), a compilation that combines a revised version of
      the LDS (LDS2) with the stray radiation corrected version of the
      Southern sky survey of the Instituto Argentino de
      Radioastronom\'ia (IAR). It is demonstrated that the selection
      criteria for dividing the parameter space are to a great extent
      independent of the particular survey in use. The situation is
      more obscure for very wide components. In this region the
      distributions of the components of different origin seem to be
      more blended and it is harder to decide on the basis of Gaussian
      parameters alone, whether the corresponding components are caused
      by some high velocity dispersion halo gas in the Milky Way,
      external galaxies or are due to baseline problems, for example.
      Nevertheless, the presence of the baseline problems in the LDS is
      most likely indicated by the peculiarities of the distribution of
      the widest Gaussians in the sky. A similar plot for the northern
      part of the LAB demonstrates considerably lower numbers of
      spurious components, but there are still problems with the
      southern part of the LAB. The strange characteristics of the
      observational noise in the southern part of the LAB are also
      pointed out.
   \end{summary}


   \resthead{Gaussian decomposition of the \ion{H}{i} surveys}{U.~Haud,
      P.~M.~W.~Kalberla}

   \newpage

   \sectionb{1}{INTRODUCTION}

      The Gaussian analysis of the observed \ion{H}{i} profiles is a
      somewhat controversial process. On the one hand, if we neglect
      the saturation and absorption effects, suppose that most of the
      line shape comes from global rotation characteristics of the
      Galaxy and that the galactic \ion{H}{i} consists of separate
      hydrogen clouds with equilibrium random velocity distribution,
      the observed profile can be considered as a sum of Gaussian cloud
      components, shifted relative to each other by differential
      rotation.

      Unfortunately, the actual situation is much more complicated. We
      must consider the possibility of intrinsically non-Gaussian
      contributions to the emission (due to groups of atoms with
      asymmetrical or in any other way pronounced non-Gaussian velocity
      distribution or due to saturation in optically thick regions and
      self-absorption by very cold foreground gas). Except for the
      simplest profiles, the least squares Gaussian analysis is not
      unique (often several quite different solutions may fit the
      observations almost equally well, and the method of the least
      squares provides no satisfactory means for choosing between these
      solutions, while other, equally good or even better ones, may not
      be found at all). Strictly speaking, the pure method of the least
      squares is even not valid, when applied to this problem, as
      neither the form of the components nor their number is known, nor
      can it be assumed that the residuals are randomly distributed.
      The solution is often partially determined by the number of
      components introduced, the initial estimates of their parameters
      and only partially by the observed profile. All this makes a
      rigorous Gaussian analysis somewhat illusive.

      These weaknesses of the Gaussian analysis were understood rather
      early (Kaper et al. 1966; Takakubo \& van Woerden 1966).
      Nevertheless, the method continued to be used up to the present
      time (e.g. Cappa de Nicolau \& P\"oppel 1986; P\"oppel et al.
      1994; Verschuur \& Peratt 1999; Verschuur 2004). This indicates
      that besides weaknesses the method must have also some benefits
      (some aspects briefly discussed in Sec. 2.2 and in Haud 2000,
      hereafter Paper I). Proceeding from this, we have created a new
      fully automatic specialized computer program for the
      decomposition of large 21-cm \ion{H}{i} line surveys into
      Gaussian components. This program has been described in Paper I
      and it represents the profiles as formal sums of only positive
      Gaussian functions without considering the actual line formation
      processes. During the decomposition process the special attention
      is paid to the following features:
      \newcounter{loend1}
      \begin{list}{\arabic{loend1}.}
         {\usecounter{loend1}
         \setlength{\parsep}{0pt}
         \setlength{\itemsep}{0pt}
         \setlength{\topsep}{0pt}}
         \item{Several quite different solutions may often fit the
            observations almost equally well. To choose from these
            solutions, it was supposed that general properties of the
            hydrogen distribution are somewhat correlated at
            neighboring sky positions and therefore the program tries
            to find similar decompositions for corresponding
            profiles.}
         \item{With the increasing complexity of the observed profiles,
            the number of Gaussians in decompositions usually grows
            rapidly and the values of their parameters become mutually
            dependent. To reduce this problem, special means have been
            used to keep the number of Gaussians as small as
            possible.}
      \end{list}

      In this paper we describe the first results from the application
      of the new decomposition program on the Leiden/Dwingeloo Survey
      of galactic neutral hydrogen by Hartmann \& Burton (1997,
      hereafter the LDS) and on the recent Leiden/Argentine/Bonn
      compilation of galactic \ion{H}{i} results by Kalberla et al.
      (2005, hereafter the LAB). We start with the description of the
      data used, the results obtained and possible approaches to the
      interpretation of the decomposition results (Section 2) and then
      proceed with the analysis of the results for the LDS (Section 3)
      and the LAB (Section 4). In this analysis, the main attention is
      paid to the separation of the features, most likely corresponding
      to the real emission of galactic \ion{H}{i}, from those
      representing different problems during observations, reduction
      and decomposition of the surveys. Such a separation may be
      important at least in two aspects. On the one hand, such
      separation may help us to identify the problems with the
      observational data; on the other hand, it may give us a cleaner
      sample of Gaussians for studying the properties of the Milky Way
      \ion{H}{i}. At the same time, it is important to understand that
      the discrimination between different Gaussians, as described in
      this paper, is statistical in its nature and as the distributions
      of different types of Gaussians partially overlap, it cannot be
      used as the only and final criterion for every particular
      Gaussian or profile.

      In the present paper, we turn our main attention to the minority
      of the Gaussians, most likely describing different types of
      problems, which could be searched for using the components,
      recognized as probably suspicious. In the further papers we will
      continue the study of the majority of the Gaussians, which may
      with higher probability measure the spectral signatures of higher
      astronomical interest.

   \sectionb{2}{INTERPRETATION OF THE DECOMPOSITION RESULTS}

      \subsectionb{2.1}{The data}

      As test data for our decomposition program we used the original
      observed profiles of the LDS (Hartmann 1994), reduced to
      $T_\mathrm{b}$ by P. M. W. Kalberla at Bonn University (the
      reduction procedures described in Hartmann 1994 and Hartmann et
      al. 1996). These are not exactly the same as those published by
      Hartmann \& Burton (1997) on a CD-ROM. We have used the profiles
      before averaging the repeated observations at identical sky
      positions and before re-gridding them onto a common lattice. This
      choice was made, as averaging and re-gridding smear the
      differences between neighboring profiles and may have undesirable
      influence on the Gaussian decomposition process (see Introduction
      of Paper I). In this original form the survey contained 184\,698
      profiles, which after decomposition were represented by
      1\,493\,187 Gaussians. For a brief comparison also 206\,671
      profiles of the published version of the LDS were decomposed into
      1\,644\,665 Gaussians.

      Recently a similar Southern sky high sensitivity \ion{H}{i}
      survey at $\delta \le -25\degr$ was published by Bajaja et al.
      (2005) (IAR). IAR and the LDS with the revised
      stray-radiation and baseline corrections (LDS2) make up the LAB.
      The specifications for LDS2 and IAR closely match each other,
      but all the data reduction and calibration procedures were
      carried out entirely independently for both of the surveys.
      Proceeding from this, also the Gaussian decomposition was carried
      out separately for the LDS2 and the IAR. Once again, the
      original data were used for both surveys. As for the LDS2 the
      repeated observations were not averaged any more, but the final
      profiles were selected on the basis of the best agreement of
      their Gaussian decompositions with the decompositions of the
      neighboring profiles, we used these preselected profiles for the
      analysis. In the case of the IAR, we first used for the
      decomposition the original 1008-channel data of all observed
      profiles and for repeated observations we applied the same
      selection criteria as used for the LDS2. This procedure gave us
      1\,064\,808 Gaussians per 138\,830 profiles for the LDS2 and
      444\,573 Gaussians per 50\,980 profiles in the case of the IAR.

      \vskip 6mm
      \subsectionb{2.2}{The usage of Gaussians}

      In general, the results of these Gaussian decompositions may be
      interpreted from two completely different points of view:
      \newcounter{loend2}
      \begin{list}{\arabic{loend2}.}
         {\usecounter{loend2}
         \setlength{\parsep}{0pt}
         \setlength{\itemsep}{0pt}
         \setlength{\topsep}{0pt}}
         \item{Gaussian parameters may be considered just as a compact
            means for representing the observed data without providing
            any physical interpretation to these parameters, or}
         \item{we may want to derive from the parameters of the
            obtained Gaussians direct information regarding the
            structure of the interstellar medium.}
      \end{list}

      For the first purpose, even in the case of the most complicated
      profiles, hundreds of channel values in the profile are replaced
      by some tens of Gaussian components, while physically significant
      information, such as mean velocities and the \ion{H}{i} content
      of the \ion{H}{i} features, can still be as easily extracted as
      from full profile data (Shane 1971). If correctly performed, the
      decomposition just discards observational noise and in this case
      we must consider all obtained Gaussians as real as the original
      profile data and the criteria for performing the decomposition.
      Moreover, usually some specific features in the observed profiles
      (not necessarily corresponding to some distinct gas clouds in
      physical space) are represented by some specific set of
      Gaussians, which can be found from the overall data-set more
      easily than un-parameterized spectral features. As noted by
      Verschuur \& Peratt (1999), such Gaussian analysis allows us to
      characterize general properties of the profiles from region to
      region in the sky and to draw conclusions, based upon
      similarities and differences in profile shapes. Verschuur has used
      this approach to study the relations between the different
      line-width regimes of the \ion{H}{i} in the local interstellar
      medium and the critical ionization phenomenon (Verschuur \&
      Schmelz 1989; Verschuur \& Magnani 1994; Verschuur \& Peratt
      1999; Verschuur 2004). This is also the way of interpretation
      used in the present paper. We are looking for patterns in the
      distribution of Gaussians, corresponding to different problems in
      the obtained decomposition.

      The second approach is much more complicated as it is well known
      that Gaussians may yield direct information regarding the
      structure of the interstellar medium only for the simplest
      profiles, where at least some Gaussians are well separated. This
      considerably reduces the usefulness of the Gaussian decomposition
      in directions close to the galactic equator, where the line of
      sight may contain hundreds of times more gas than at higher
      latitudes. The situation could not be improved even by increasing
      the resolution of observations, as the shape of the emission
      spectra does not change greatly with angular resolution (Baker \&
      Burton 1979). This is so because near the galactic plane the more
      or less continuous distribution of \ion{H}{i} extends to
      distances much larger than at high latitudes and the velocity
      crowding squashes a lot of space into a few kilometers per
      second.  Individual interstellar components blend completely, and
      little structure is revealed by increasing the resolution. Only
      when an emission feature has a very odd velocity or is
      considerably brighter than its surroundings, so that it is not
      blended by other emission, the Gaussian analysis may still
      provide some information on the properties of the interstellar
      gas (limits on velocity dispersion and temperature, for example).
      Due to these considerations, we first use for our discussions
      only the decomposition results obtained for relatively high
      galactic latitudes $(|b| \ge 30\degr)$, but afterwards we test
      also how much and in what general properties of the decomposition
      results differ for the regions near the galactic plane and
      farther away.

      However, even if we consider only relatively simple profiles at
      high galactic latitudes, some of the well separated Gaussians,
      obtained during the decomposition process, may correspond to
      other features in profiles than the real emission from galactic
      \ion{H}{i}. Therefore, when in an ideal case, the Gaussian
      analysis can be used to separate the useful signal from the
      observational noise, in reality such a separation is never
      perfect. For example, some Gaussians obtained may still represent
      the inaccuracies in the termination criteria of the decomposition
      process, when stronger noise peaks are still fitted by Gaussians
      (actually, for our program these criteria were deliberately
      chosen in the way preferring fitting of some noise to loosing a
      weak signal~-- see Paper I), or suspicious features
      (radio-interferences, bad baseline and so on) contained in
      profiles themselves. Proceeding from this, the main focus of this
      paper is on the question, whether it is somehow possible to
      separate such Gaussians from other, physically more founded ones,
      and in this way to further clean the decomposition results. As a
      working hypothesis, we expect that the distributions of the
      parameter values of the Gaussians, representing different
      artifacts, may be distinguishable from those of the Gaussians
      corresponding to the real \ion{H}{i} emission of the Milky Way
      and it may be possible to isolate the regions of the parameter
      space dominated by components of one or another origin.

   \sectionb{3}{LDS}

      \subsectionb{3.1}{Narrow Gaussians}

      \begin{wrapfigure}[20]{r}[0pt]{75mm}
         \vskip -6mm
         \psfig{figure=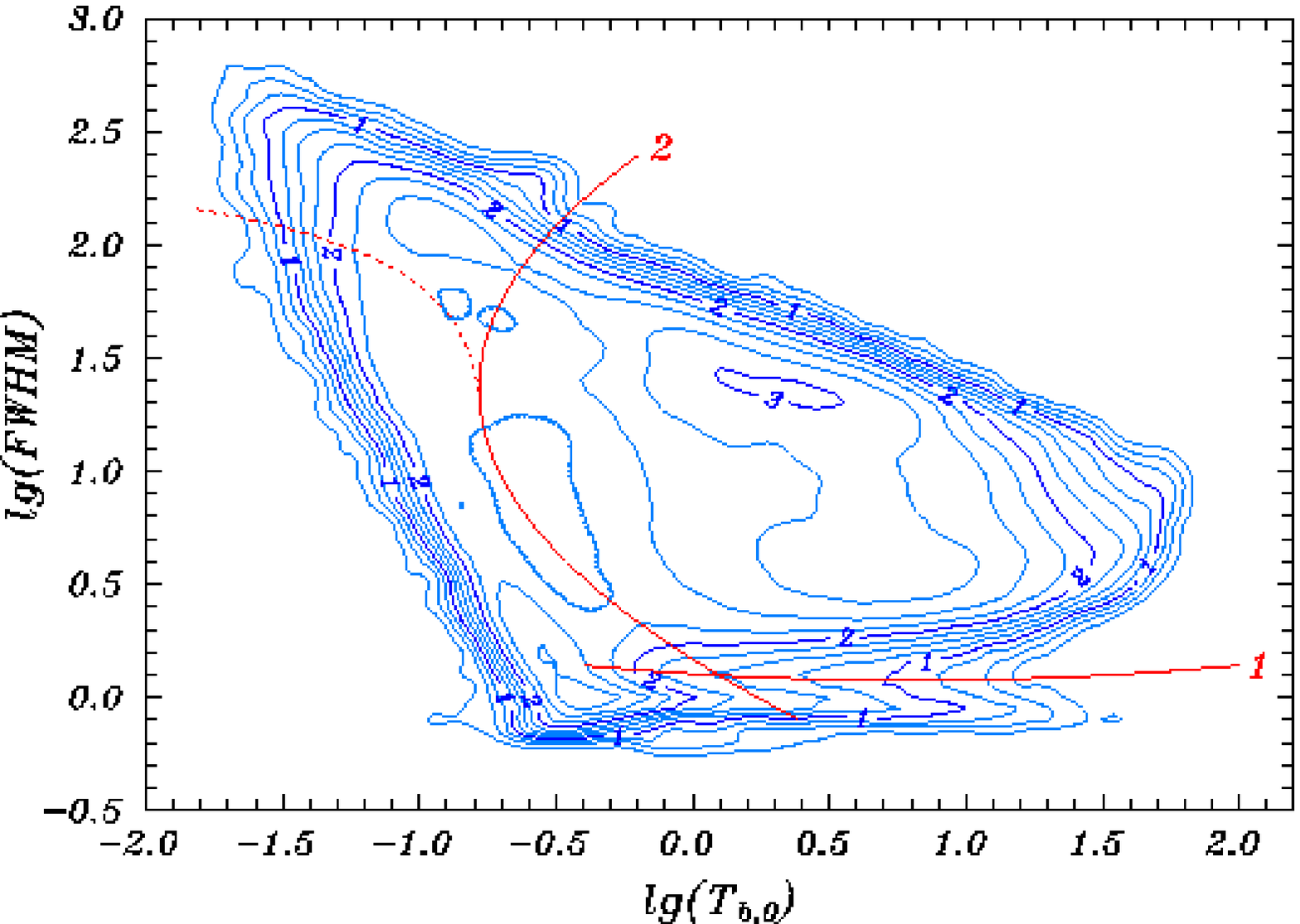,width=73mm,clip=}
         \vskip 1mm
         \captionb{1}{Frequency distribution of the parameter values in
            $(\lg(T_\mathrm{b0}), \lg(\mathrm{FWHM}))$ plane for all
            Gaussians, corresponding to profiles at galactic latitudes
            $|b| \ge 30\degr$. Isodensity lines are drawn in the scale
            of $\lg(N+1)$ with the interval of 0.25. The thick solid
            and dashed red lines represent the selection criteria
            discussed in the text.}
      \end{wrapfigure}
      For separation of different kinds of Gaussians their distribution
      in the plane of height and width seems to be most informative.
      The height of a Gaussian is defined by the value of the central
      brightness temperature $T_\mathrm{b0} > 0$ from the standard
      Gaussian formula
      \begin{equation}
         T_\mathrm{b} = T_\mathrm{b0}
                   \mathrm{e}^{-\frac{(V - V_C)^2}
                   {2\sigma_V^2}}, \label{Eq1}
      \end{equation}
      where $T_\mathrm{b}$ is the brightness temperature and $V$ is the
      velocity of the gas relative to the Local Standard of Rest. $V_C$
      is the velocity corresponding to the center of the Gaussian. We
      characterize the widths of the components by their full width at
      the level of half maximum ($\mathrm{FWHM}$), which is related to
      the velocity
      dispersion $\sigma_V$ by a simple scaling relation $\mathrm{FWHM}
      = \sqrt{8 \ln{2}} \sigma_V$.

      To avoid the complicated profiles near the galactic plane, we
      present in Fig.~1 the $(\lg(T_\mathrm{b0}),~\lg(\mathrm{FWHM}))$
      distribution of all Gaussians for the LDS profiles at galactic
      latitudes $|b| \ge 30\degr$. From this figure we can see that
      most frequently the Gaussians have the heights between about $1
      \la T_\mathrm{b0} \la 10~\mathrm{K}$ and widths $3 \la
      \mathrm{FWHM} \la 30~\mathrm{km\,s}^{-1}$. These parameters are
      in general agreement with the usual two-phase models of the
      atomic interstellar medium, where one phase is cold with
      temperatures of about $100~\mathrm{K}$ (CNM) and the other is
      warm with temperatures of several thousands degrees (WNM).
      Considering also the temperature variations and the additional
      line broadening due to the macroscopic turbulent motions in the
      interstellar structures ($\sigma_V \simeq 2 -
      5~\mathrm{km\,s}^{-1}$ according to Burton 1992), the widths of
      corresponding \ion{H}{i} emission lines are in the range of about
      $1 \la \mathrm{FWHM} \la 10~\mathrm{km\,s}^{-1}$
      \begin{wrapfigure}[36]{r}[0pt]{75mm}
         \psfig{figure=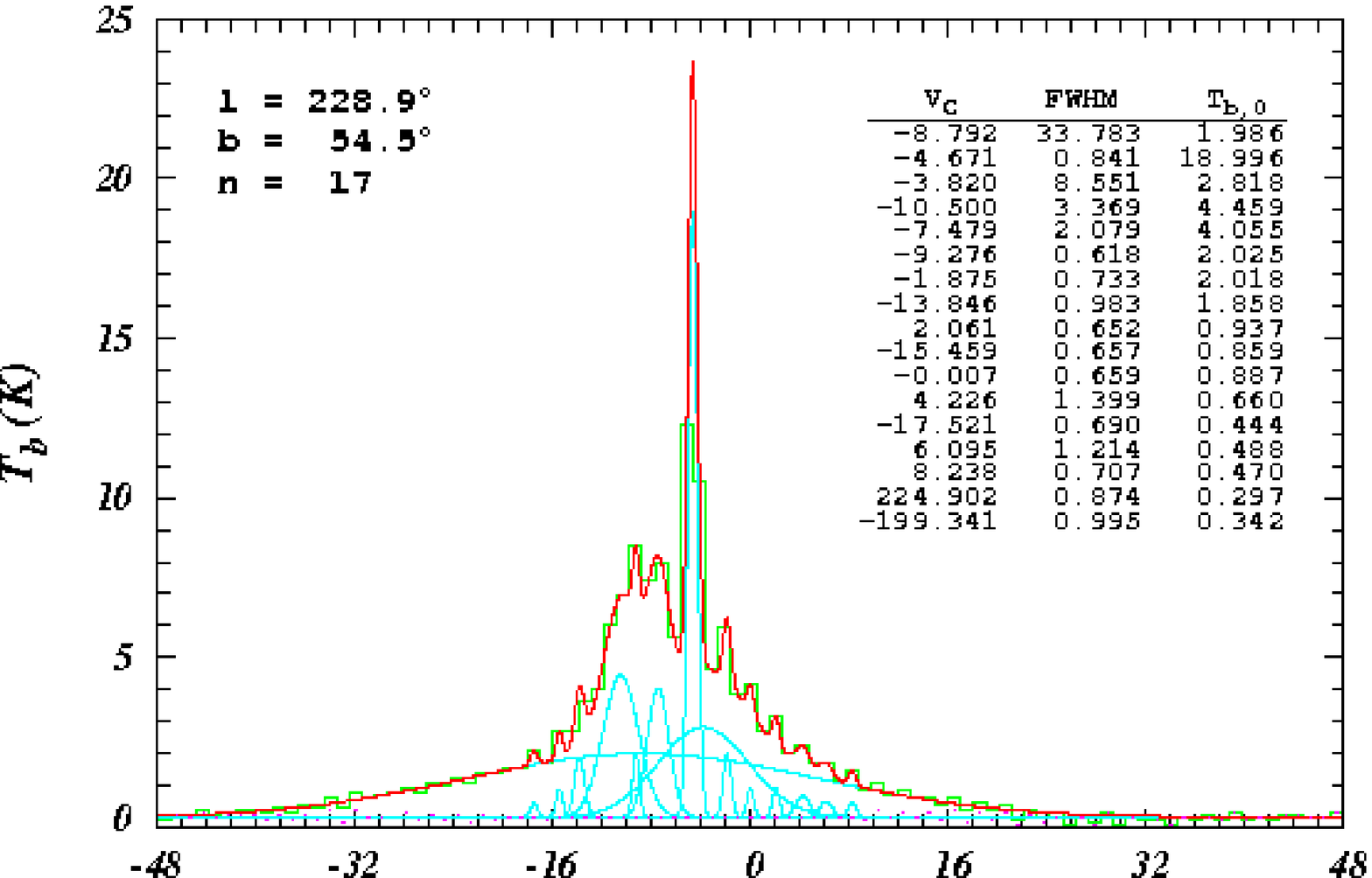,width=73mm,clip=}
         \vskip 3mm
         \psfig{figure=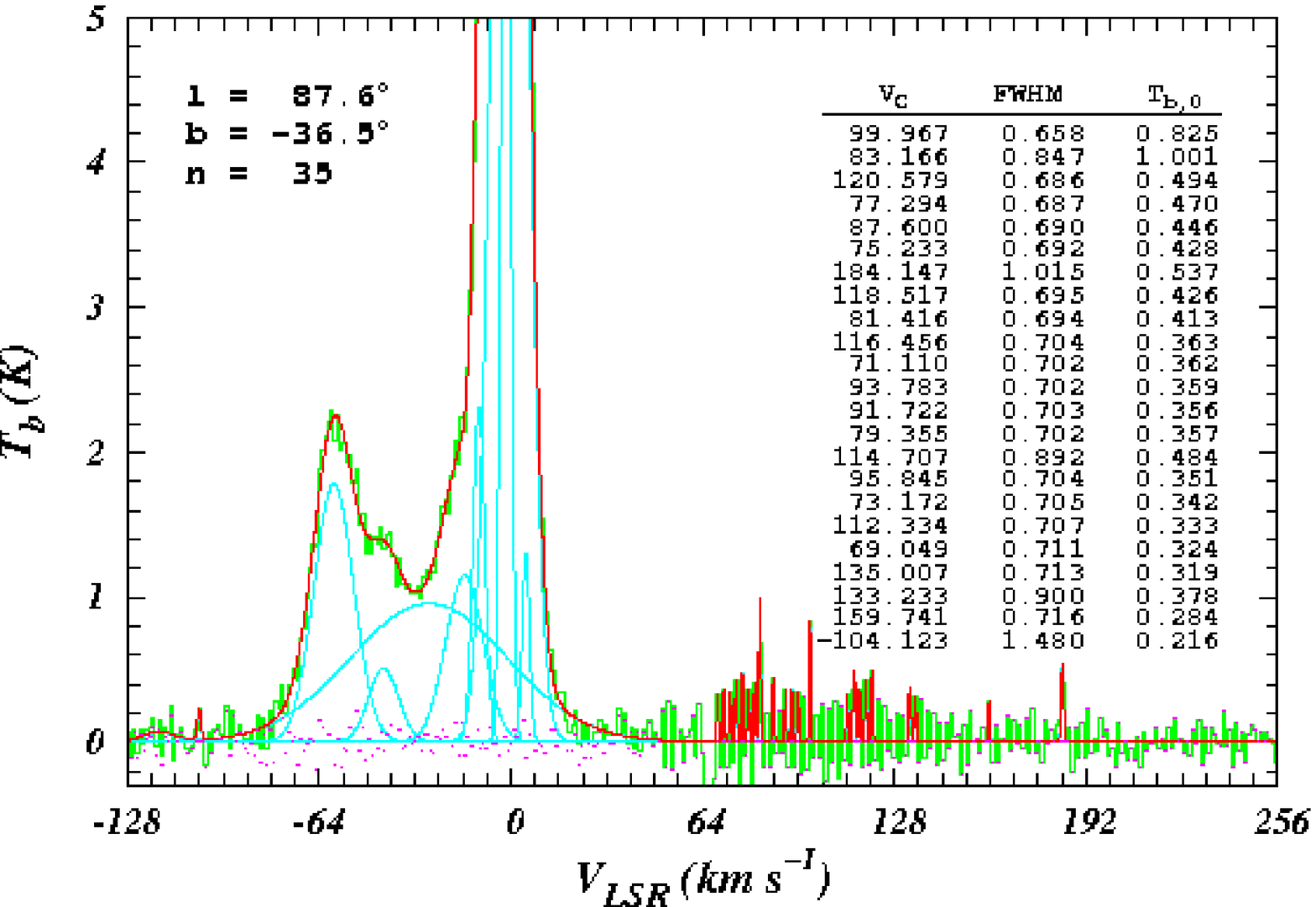,width=73mm,clip=}
         \vskip 1mm
         \captionb{2}{Two examples of the profile with very narrow
            Gaussians. The observed profiles are plotted with the
            stepped green lines, individual Gaussian components with
            the thin smooth cyan lines, the Gaussian representation of
            the profiles with the thick smooth red lines and the
            residuals with small magenta points. Due to the large
            number of Gaussians in lower panel only the parameters of
            very narrow components are given.}
      \end{wrapfigure}
      for CNM
      (Crovisier 1981) and $12 \la \mathrm{FWHM} \la
      40~\mathrm{km\,s}^{-1}$ for WNM (Mebold 1972).  There are,
      however, concentrations of Gaussians also around
      $\lg(T_\mathrm{b0}) \approx -0.4$, $\lg(\mathrm{FWHM}) \approx 0$
      and $\lg(T_\mathrm{b0}) \approx -0.9$, $\lg(\mathrm{FWHM})
      \approx 2.1$.

      The first of these concentrations extends from relatively weak
      Gaussians up to the intensities of several tens of kelvins, and
      the widths are mostly below the limit corresponding to the
      kinetic temperature of the coldest \ion{H}{i} observed (Verschuur
      \& Knapp 1971, 1972; Braun \& Burton 2000). Therefore, it is
      likely that these components are not directly related to the
      emission of the galactic gas, but represent some artifacts of the
      profiles or their Gaussian decomposition process. An inspection
      of corresponding profiles confirms this guess. In Fig.~2 two
      examples are given. The upper panel represents a \ion{H}{i}
      profile, measured at $l = 228.9\degr$, $b = 54.5\degr$ and
      decomposed into 17 Gaussians of which 13 have widths around
      $1~\mathrm{km\,s}^{-1}$. One of these narrow Gaussians is rather
      strong with the height of nearly $19~\mathrm{K}$, representing
      the central peak of the typical radio-interference, and most of
      the other Gaussians fit the fading oscillations on both sides of
      the central peak. There are also two narrow Gaussians (not shown
      in Fig. 2) at higher velocities, which represent some stronger
      noise peaks, not directly related to the illustrated interference
      pattern. The lower panel illustrates the Gaussians fitting the
      noise peaks. As the noise level at the velocities around
      $100~\mathrm{km\,s}^{-1}$ is higher (probably due to the
      corrected radio-interference) than in other regions of the
      profile, this higher noise have brought about a large number of
      spurious Gaussians.

      As the described very narrow Gaussians are separated in Fig.~1
      from the distribution of the wider ones by a clearly visible
      ``valley'' of relatively underrepresented values of Gaussian
      parameters, it seems rather safe to draw the separation line
      between the Gaussians describing the emission of the galactic
      \ion{H}{i} and those representing the noise and
      radio-interferences, along the bottom of this valley (line 1 in
      Fig.~1). To select the shape and placement of this line, we first
      counted the Gaussians in $(0.05 \times 0.05)$ bins in
      $\lg(T_\mathrm{b0})$ and $\lg(\mathrm{FWHM})$, found for every
      strip of $-0.3 \le \lg(T_\mathrm{b0}) \le 1.2$ the value of
      $\lg(\mathrm{FWHM})$, corresponding to the bin with the smallest
      number of Gaussians and fitted the parabola through the obtained
      $(\lg(T_\mathrm{b0}),~\lg(\mathrm{FWHM}))$ pairs. Finally, the
      parameters of the parabola were adjusted by demanding that the
      sum of the numbers of Gaussians in the bins, through which the
      parabola is drawn, should be minimal. In some approximation the
      result,
      \begin{equation}
         \lg(\mathrm{FWHM}) = 0.046*\lg(T_\mathrm{b0})^2 -
                              0.074*\lg(T_\mathrm{b0}) +
                              0.104, \label{Eq2}
      \end{equation}
      could be used as a selection criterion to separate the Gaussians
      representing interferences and noise peaks from those describing
      the properties of the emission of the galactic \ion{H}{i}.

      \subsectionb{3.2}{Weak Gaussians of intermediate widths}

      \begin{wrapfigure}[18]{r}[0pt]{75mm}
         \vskip -4mm
         \psfig{figure=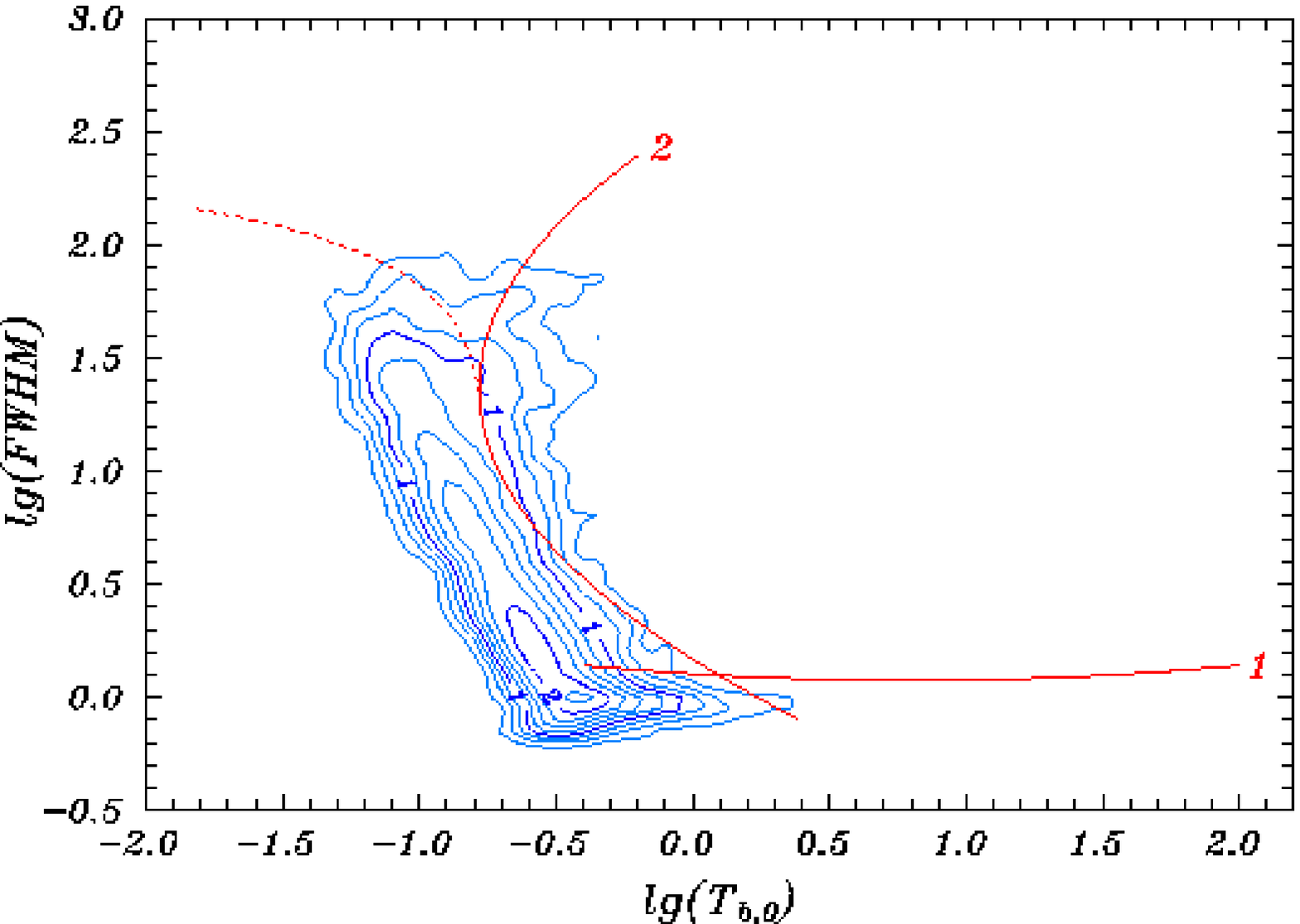,width=73mm,clip=}
         \vskip 1mm
         \captionb{3}{The frequency distribution of the parameters of
            the Gaussians near the profile edges. The designation is
            the same as in Fig.~1.}
      \end{wrapfigure}
      Around $\lg (T_\mathrm{b0}) \approx 0$ and $\lg (\mathrm{FWHM})
      \approx 0$ the valley in Fig.~1 turns towards the wider
      Gaussians, broadens and becomes less deep. Therefore, it is
      interesting to check, if here also the region of underrepresented
      values of Gaussian parameters may help us to separate the
      components describing different phenomena. The answer seems to be
      yes. It is known that the receiver bandbass is never square. As a
      result, the extreme edges of the obtained spectra are steeply
      falling off to zero intensity and after the bandpass removal the
      intensities in these channels become unreliable due to the
      division by reference. Moreover, the usual methods of baseline
      fitting are poorly constrained in these regions. To take this
      into account, during the decomposition we have not used the data
      from 64 channels with the most negative velocities and from 128
      channels with the highest positive velocities. However, the
      excess noise and baseline problems at the edges of the reduced
      profiles are not confined by sharply defined velocity ranges. In
      some profiles they can be detected in wider regions, in other
      profiles in more limited regions. To study how this is expressed
      in the ``language'' of Gaussians, we plot in Fig.~3 the
      distribution of the Gaussians, which have central velocities
      closer to the accepted profile ends than 64 channels. We can see
      that these Gaussians populate the region separated in Fig.~1 from
      the main body of the distribution by the valley running towards
      the upper left-hand corner of the plot.

      \begin{wrapfigure}[29]{r}[0pt]{75mm}
         \vskip -4mm
         \psfig{figure=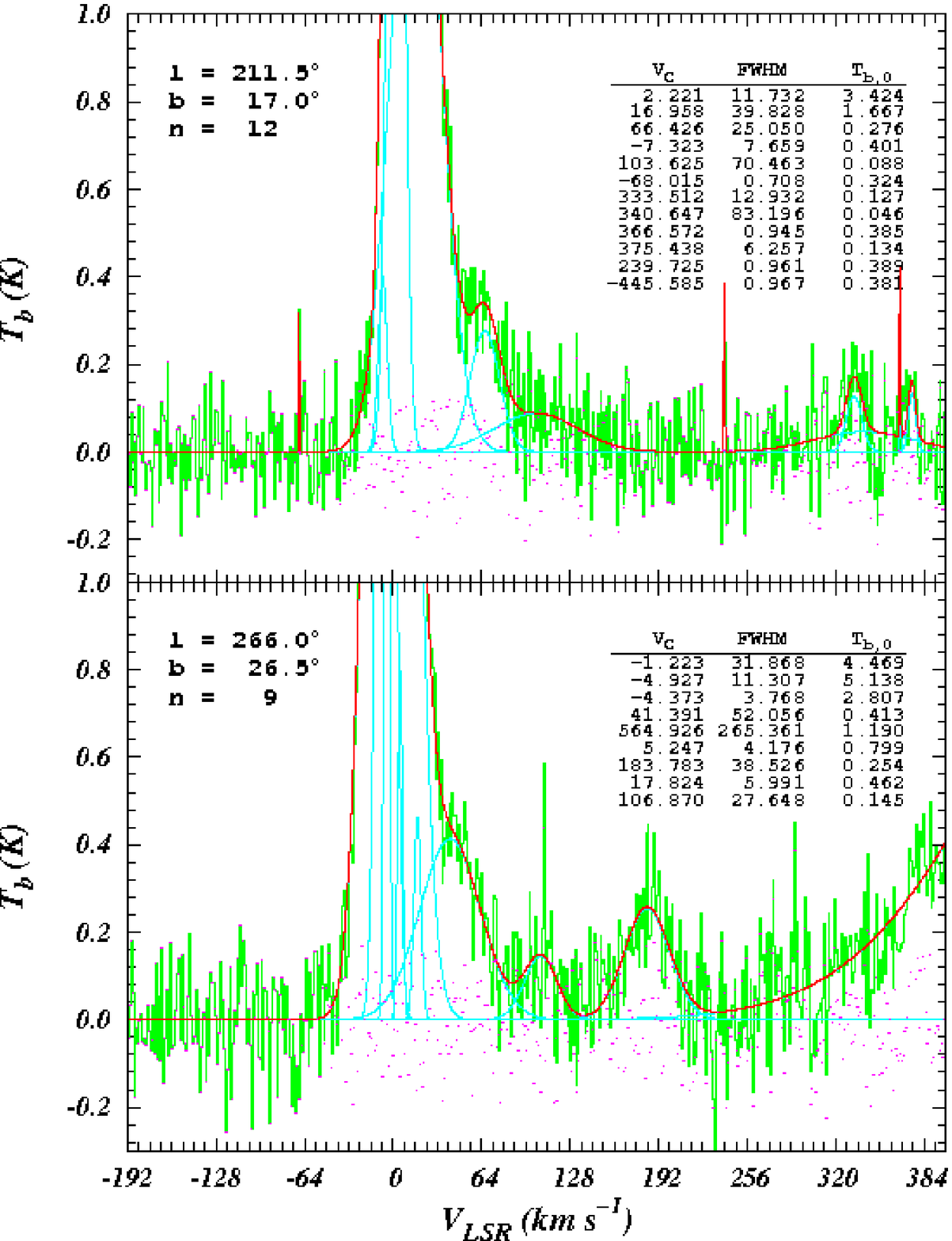,width=73mm,clip=}
         \vskip 1mm
         \captionb{4}{Two examples of the profiles with the problems at
            the positive velocity edge. The notation is the same as in
            Fig.~2.}
      \end{wrapfigure}
      In Fig.~4 two examples are given. The upper panel gives the
      \ion{H}{i} profile, measured at $l = 211.5\degr$, $b = 17.0\degr$
      and decomposed into 12 Gaussians, 4 of which have central
      velocities higher than $330~\mathrm{km\,s}^{-1}$~-- the velocity
      limit, corresponding to the 192th channel from the positive
      velocity edge of the observed profile. We can see that at these
      extreme velocities the baseline has been drawn too low, causing
      the resulting mean profile to arise above the zero intensity
      level. In the Gaussian decomposition this behavior is expressed
      by adding a rather weak ($T_\mathrm{b0} \approx 0.05~\mathrm{K}$)
      but broad component with $\mathrm{FWHM} \approx
      83~\mathrm{km\,s}^{-1}$. The profile also becomes wavy in this
      region. As a result, the decomposition program adds two more weak
      Gaussians at velocities $V_C \approx 333~\mathrm{km\,s}^{-1}$ and
      $375~\mathrm{km\,s}^{-1}$. Finally, as the program has considered
      the increase of the profile mean channel values at extreme
      velocities as a sign of presence of a possible useful signal, the
      actual noise level of the profile is
      underestimated. As the
      decomposition program tries to reduce the level of residuals to
      the value near the estimated noise level of the profile, this
      results in several very narrow Gaussians, representing the
      strongest noise peaks spread all over the profile.

      The narrow Gaussians in Fig.~4 could be detected and rejected
      from the obtained Gaussian representation of the LDS using the
      selection criterion 1 (Eq. \ref{Eq2}) discussed above. As
      demonstrated by Fig.~3, the wider Gaussians near the profile
      edges in Fig.~4 fall to the left of the region of
      underrepresented values of the Gaussian parameters in Fig.~1.
      Therefore, it seems plausible that even after the valley in the
      distribution of the Gaussian parameters in Fig.~1, turns up near
      $\lg (T_\mathrm{b0}) \approx 0$ and $\lg (\mathrm{FWHM}) \approx
      0$, it still may be interpreted as the division between
      Gaussians, representing the actual emission of the galactic
      \ion{H}{i} and those describing the observational, reductional
      and decompositional problems. On the basis of these
      considerations, we followed the underpopulated region to even
      wider Gaussians and approximated the run of the location of the
      most sparsely populated parameter values with the parabolic curve
      (solid line 2 in Figs.~1 and 3).

      However, in Fig.~3 we can also see that in the region of the
      widest Gaussians a considerable fraction of them fall to the
      right of the parabolic curve, determined from the location of the
      distribution minimum in Fig.~1. A closer inspection of the
      situation indicates that the majority of these Gaussians have
      central velocities outside the range of channels, actually used
      in the decomposition process. That means, they are similar to the
      $\mathrm{FWHM} \approx 83~\mathrm{km\,s}^{-1}$ Gaussian in
      Fig.~4, but with its center not at $V_C \approx
      340~\mathrm{km\,s}^{-1}$, but shifted beyond the right border of
      the figure. These Gaussians arise in the cases where the baseline
      has been drawn progressively lower and lower towards the end of
      the used velocity range and the resulting profile continues to
      rise higher and higher above the zero level when approaching the
      velocity limit of the profile, as illustrated in the lower panel
      of Fig.~4. The parameters of such Gaussians are poorly determined
      as they are estimated from the relatively small number of profile
      channels, covering less than a half of the full extent of the
      component. These Gaussians could be easily recognized by
      demanding that all the accepted components must have their
      central velocities inside the velocity range, used for
      decomposition.

      \subsectionb{3.3}{The broadest Gaussians}

      We have determined the shape and location of curve 2 in Fig.~3 in
      the same way as described for curve 1. However, when at lower
      values of $\mathrm{FWHM}$ curve 2 can be rather easily determined
      from the data used for Fig.~1, this becomes increasingly
      uncertain when moving to higher values of $\mathrm{FWHM}$. Above
      $\lg (\mathrm{FWHM}) \approx 1.35$ the valley bifurcates and the
      curve may actually follow even two completely different passages:
      the one indicated in Fig.~1 with the solid line, or the other
      one, indicated by the dashed line. The main difference between
      these two possibilities is that in the first case the widest
      Gaussians, obtained during the decomposition, are excluded from
      the region corresponding to the components representing the
      actual emission of the galactic \ion{H}{i}, and in the second
      case, they are included in this region. Therefore, it is
      important to know what is actually represented by such Gaussians.

      Some decades ago, Field et al. (1969) demonstrated that the
      \ion{H}{i} could be considered as a two-phase medium,
      where much of the gas is observed to be either WNM with $T
      \sim 10^4~\mathrm{K}$ or CNM with $T \sim 100~\mathrm{K}$
      (Kulkarni \& Heiles 1987; Dickey \& Lockman 1990). These
      temperatures correspond to line-widths mostly below
      $21~\mathrm{km\,s}^{-1}$ and the corresponding Gaussians have
      $\lg (\mathrm{FWHM}) \la 1.9$ even if we allow for realistic
      turbulent motions in the gas (Mebold et al. 1982; Kulkarni \&
      Fich 1985). Kalberla et al. (1998) have
      argued in favor of the existence of some neutral gas with
      velocity dispersion as high as $60-80~\mathrm{km\,s}^{-1}$ in the
      halo. The Gaussians, corresponding to such gas may have $\lg
      (\mathrm{FWHM}) \la 2.3$, the limit still considerably below the
      highest values of $\lg (\mathrm{FWHM})$ in Fig.~1. Therefore, it
      seems clear that the widest Gaussians in Fig.~1 cannot be
      interpreted as the representation of the real \ion{H}{i} emission
      and this favors the solid line 2 as a selection criterion.

      \begin{wrapfigure}[20]{r}[0pt]{75mm}
         \psfig{figure=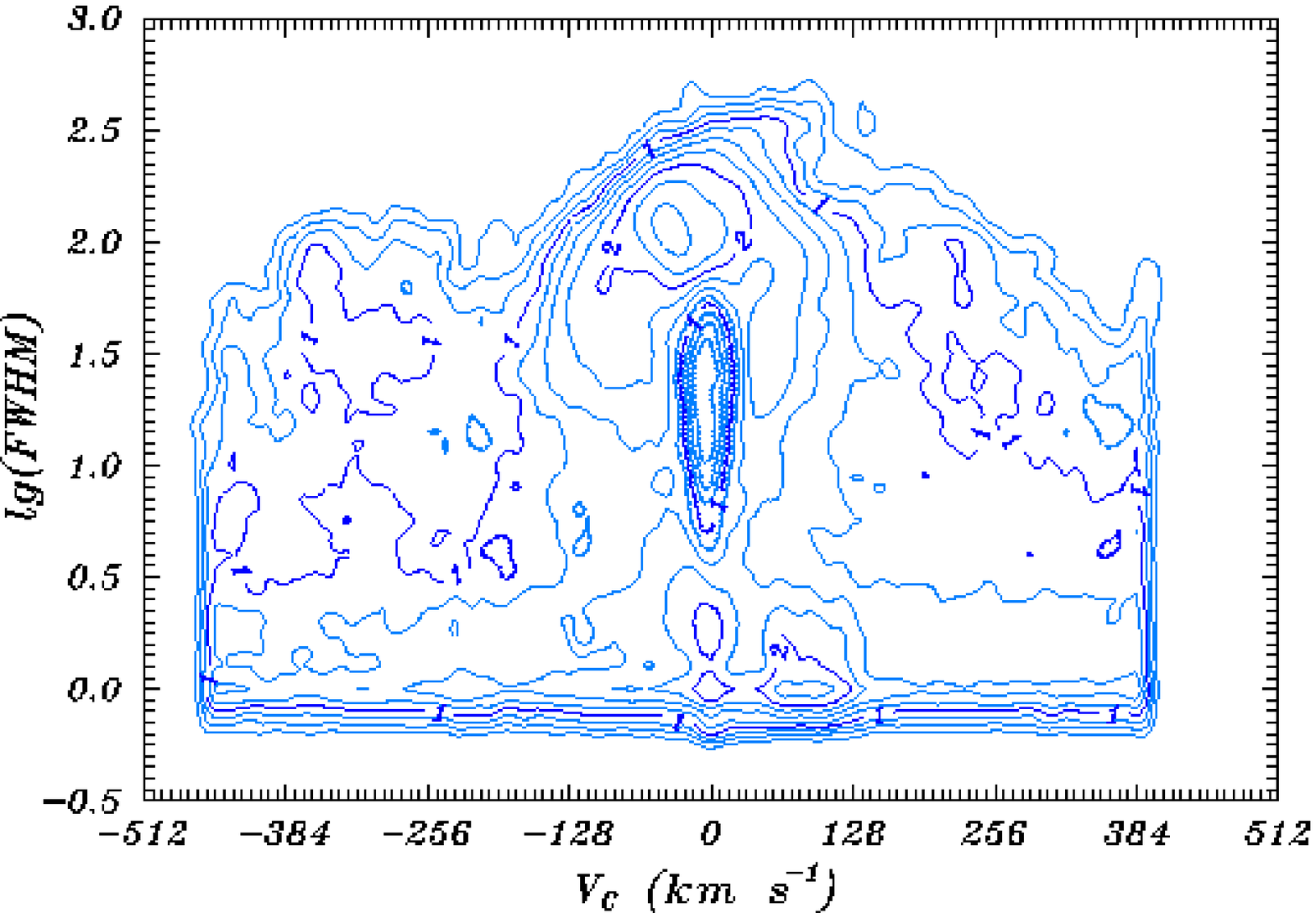,width=73mm,clip=}
         \vskip 1mm
         \captionb{5}{The distribution of the weakest Gaussians (those
            remaining to the left of the solid curve 2 in Fig.~1) in
            the central velocity~-- line-width plane. The isodensity
            lines are drawn in the scale of $\lg(N+1)$ with the
            interval of 0.25.}
      \end{wrapfigure}
      However, what is then represented by the widest Gaussians
      rejected by the second selection criterion? To discuss this, we
      first turn to the distribution of the weak Gaussians, remaining
      to the left of the solid curve 2 in Fig.~1, in the central
      velocity~-- line-width plane (Fig.~5). Here we can see that the
      widest Gaussians are concentrated in their velocities mainly
      around the value of $V_C = 0~\mathrm{km\,s}^{-1}$ with the
      distribution extending to slightly more than
      $150~\mathrm{km\,s}^{-1}$ at both sides. Less prominent
      concentrations of wide Gaussians are at higher velocities
      (approximately at $-390 \la V_C \la -230~\mathrm{km\,s}^{-1}$ and
      $180 \la V_C \la 280~\mathrm{km\,s}^{-1}$). There is a certain
      excess of wide Gaussians also near the profile edges, but we have
      already discussed them above.

      \vbox{
         \vskip 3mm
         \centerline{\psfig{figure=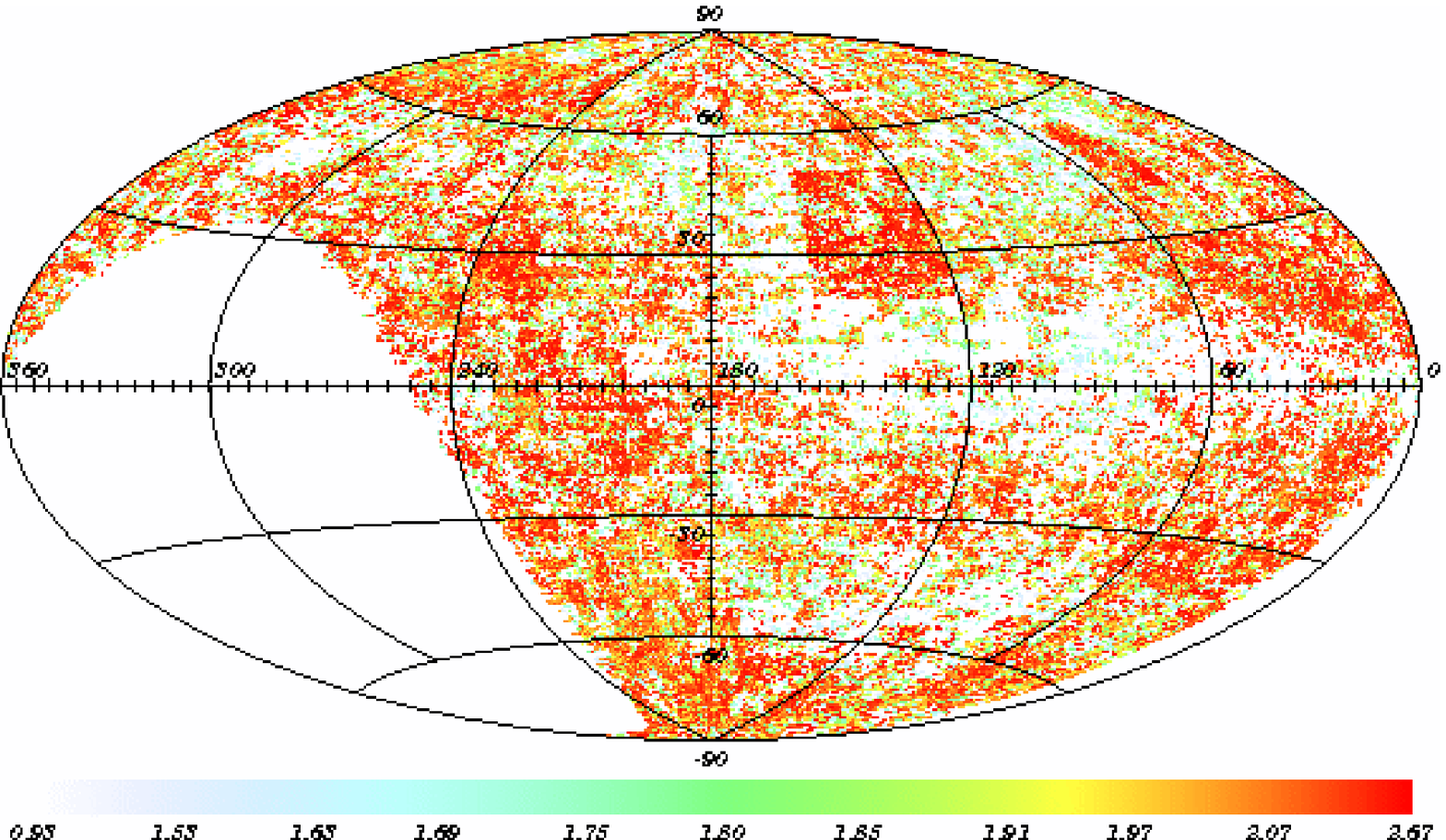,width=125mm,angle=0,clip=}}
         \vskip 1mm
         \captionb{6}{The sky distribution of the widest weak Gaussians
            in galactic coordinates. The color-scale represents the
            width (in units of $\lg (\mathrm{FWHM})$) of the widest
            Gaussian obtained for a given sky position and the
            gradation is chosen to enhance the contrast of quadrangular
            fields.}
      }
      \vspace{3mm}
      In Fig.~6 we present the sky distribution of the Gaussians of the
      concentration near the zero velocity in Fig.~5 (velocities $-152
      \la V_C \la 160~\mathrm{km\,s}^{-1}$). We can see that in
      surprisingly many places (best visible at $l > 180 \degr$ and $b
      < -30 \degr$) these components form quadrangular ``clouds'' in
      the sky (a similar pattern is not visible if we use the Gaussians
      lying to the right of the solid curve 2 in Fig.~1). The size of
      these clouds is often $5 \times 5 \degr$ or an integer multiple
      of this. If we recall now that the LDS observations were made by
      $5 \times 5 \degr$ fields and the same fields were involved also
      in bandpass removal (Hartmann 1994), it seems that at least a
      considerable number of wide Gaussians must be due to some
      observational or reductional problems specific to each $5 \times
      5 \degr$ field and cannot describe the actual properties of
      galactic \ion{H}{i}. This is the main justification for choosing
      the second selection criterion as indicated by the solid line 2
      in Fig.~1, corresponding to
      \begin{equation}
         \lg(T_\mathrm{b0}) = 0.547*\lg(\mathrm{FWHM})^2 -
                              1.492*\lg(\mathrm{FWHM}) +
                              0.235. \label{Eq3}
      \end{equation}

      \subsectionb{3.4}{The high velocity dispersion halo gas}

      \begin{wrapfigure}[30]{r}[0pt]{75mm}
         \vskip -4mm
         \psfig{figure=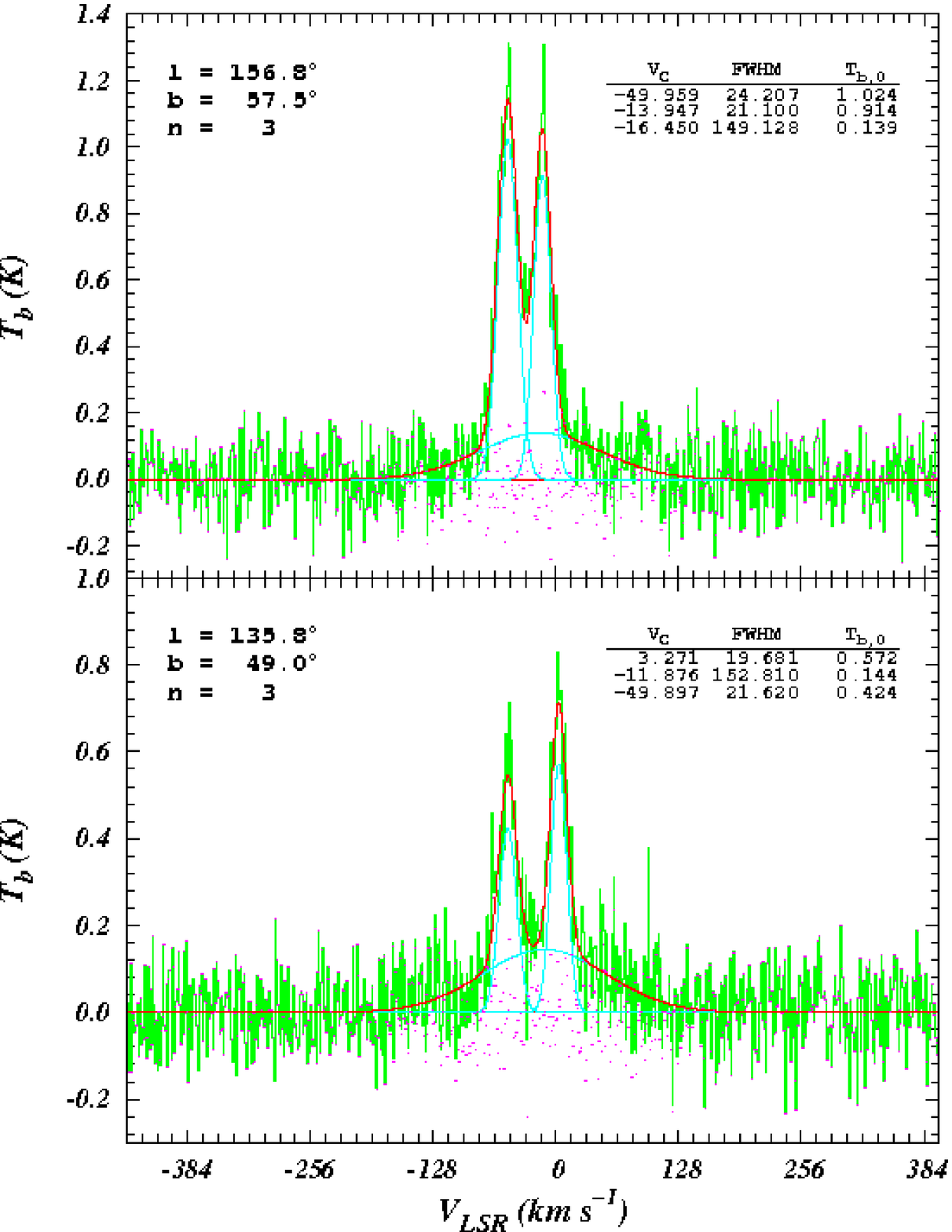,width=73mm,clip=}
         \vskip 1mm
         \captionb{7}{The examples of the profiles with broad
            Gaussians. The cases represented in the upper and lower
            panel are discussed in the text. The notation is the same
            as in Fig.~2.}
      \end{wrapfigure}
      When constructing Fig.~6, we tried several versions of the
      color-scale to find the one, which enhances most the contrast of
      the quadrangular structures. We found that this pattern is
      dominating among the Gaussians with widths above
      $\lg(\mathrm{FWHM}) \ga 1.8$, a result in good agreement with
      Fig.~5, where we can see that the concentration of very wide
      Gaussians around zero velocity extends down to the widths
      $\lg(\mathrm{FWHM}) \ga 1.8$. At the same time, if to draw a
      figure similar to Fig.~6, but for wide Gaussians at velocity
      intervals $-390 < V_C < -230~\mathrm{km\,s}^{-1}$ and $180 < V_C
      < 280~\mathrm{km\,s}^{-1}$, we cannot see the quadrangular
      pattern as in Fig.~6 and the picture is dominated by two
      concentrations of points, clearly coinciding in location and
      shape with the northern tip of the Magellanic Stream and the
      high-velocity cloud (HVC) complex AC. Therefore, it seems that
      not all wide Gaussians rejected by the selection criterion of Eq.
      (\ref{Eq3}), are due to observational or reductional problems,
      but some of them correspond to real gas.

      Moreover, at this level of discussion it remains unclear if and
      how Gaussians represent the high velocity dispersion halo gas
      (HVDHG) discussed by Kalberla et al. (1998). From their Fig.~1 we
      may estimate that corresponding Gaussians, if present in our
      decomposition, must have the heights of the order of
      $T_\mathrm{b0} \approx 0.05~\mathrm{K}$. In our Fig.~1 they lie
      to the left of the solid line 2 and would be rejected by the
      brute force application of the second selection criterion. We
      postpone the detailed discussion of this question to further
      papers, but we point out here that we believe that our
      decomposition has detected such gas. Most likely the
      corresponding Gaussians may add somewhat to the concentration
      around $V_C = -30~\mathrm{km\,s}^{-1}$, $\lg(\mathrm{FWHM}) =
      2.2$ in Fig.~5 and form part of the more diffuse background in
      Fig.~6. Nevertheless, it is clear that most of the widest
      Gaussians in Fig.~1 do not represent the properties of the real
      gas, but it is harder to draw a clear cut separation line between
      different types of Gaussians in this region than at smaller
      widths. In such a separation process we cannot rely only on
      heights and widths of Gaussians, but we must also consider at
      least their velocities.

      Actually it is as hard to classify corresponding Gaussians as to
      decide when the broad wings of the \ion{H}{i} emission lines in
      the observed profiles represent the HVDHG and when they are
      caused by some problems, most likely the badly behaving baselines
      or stray-radiation corrections. We illustrate this in Fig.~7,
      where the profile in the upper panel is selected from the light
      $5 \times 5 \degr$ field in Fig.~6 (the median width of the
      widest Gaussians of every point of the field is equal to
      $19.7~\mathrm{km\,s}^{-1}$) and the profile in the lower panel
      from the dark one (the median equals to
      $187~\mathrm{km\,s}^{-1}$). In this way we may expect that the
      profile in the lower panel probably has its widest Gaussian due
      to baseline problems and in the upper panel the widest Gaussian
      is more likely caused by HVDHG. However, the widest Gaussians in
      both panels have nearly the same widths and intensities and also
      the general shapes of the profiles seem to be rather similar.

      \begin{wrapfigure}[11]{r}[0pt]{53mm}
         \vskip -1mm
         \tabcolsep=8pt
         \begin{tabular}{rrr}
            \multicolumn{3}{c}{\parbox{48mm}{{\normbf Table 1.}
                        {\norm A four-component fit.}}}\\
            \noalign{\smallskip}
            \tablerule
            \noalign{\smallskip}
               \multicolumn{1}{c}{$V_C$} & $\mathrm{FWHM}$ & $T_\mathrm{b0}$ \\
            \noalign{\smallskip}
            \tablerule
            \noalign{\smallskip}
               $-51.994$ & $18.660$ & $0.780$ \\
               $-13.770$ & $12.198$ & $0.705$ \\
               $-28.634$ & $71.642$ & $0.517$ \\
               $ 83.636$ & $36.210$ & $0.087$ \\
            \noalign{\smallskip}
            \tablerule
         \end{tabular}
      \end{wrapfigure}
      From the observed profile in the upper panel of Fig.~7 it may
      seem that the decomposition with 4 Gaussians (with parameters
      given in Table~1) may give a better model with more realistic
      widths of Gaussians. A closer inspection does not confirm this
      expectation. Before decomposition the noise level of the
      signal-free regions of this profile was estimated to be equal to
      $0.0864~\mathrm{K}$. If we divide the profile into two regions,
      where the total intensity of the obtained Gaussians is below 10\%
      of this noise level (signal-free region) and above 10\% of the
      noise level (the region with signal), we can compute in both
      regions the $\mathrm{rms}$ of residuals after subtracting the
      obtained Gaussians from the observed profile. In the case of a
      three component fit the corresponding numbers are
      $0.0865~\mathrm{K}$ and $0.0864~\mathrm{K}$, respectively. We see
      that this model identifies nearly equal noise levels in both
      regions of the profile~-- a result we should expect (by applying
      corresponding weights we have taken into account the dependence
      of the noise on signal strength, as described in Paper I). For a
      four component fit we receive $0.0871~\mathrm{K}$ and
      $0.0790~\mathrm{K}$, clearly indicating that we have used too
      many Gaussians and made the $\mathrm{rms}$ of the residuals in
      the region containing the signal, too low.

      Therefore, on the basis of the present discussion, we cannot
      determine unambiguously the reason for the broad Gaussians. Some
      of them may correspond to real \ion{H}{i} emission, others to
      artifacts due to the problems in baseline determination. However,
      even for wide Gaussians Figs.~1, 5 and 6 may help us separate the
      regions of the parameter space, where such problems exist, from
      those, where most components with high probability describe the
      real emission.

      \subsectionb{3.5}{Low galactic latitudes and the Atlas}

      \begin{wrapfigure}[36]{r}[0pt]{75mm}
         \vskip -4mm
         \psfig{figure=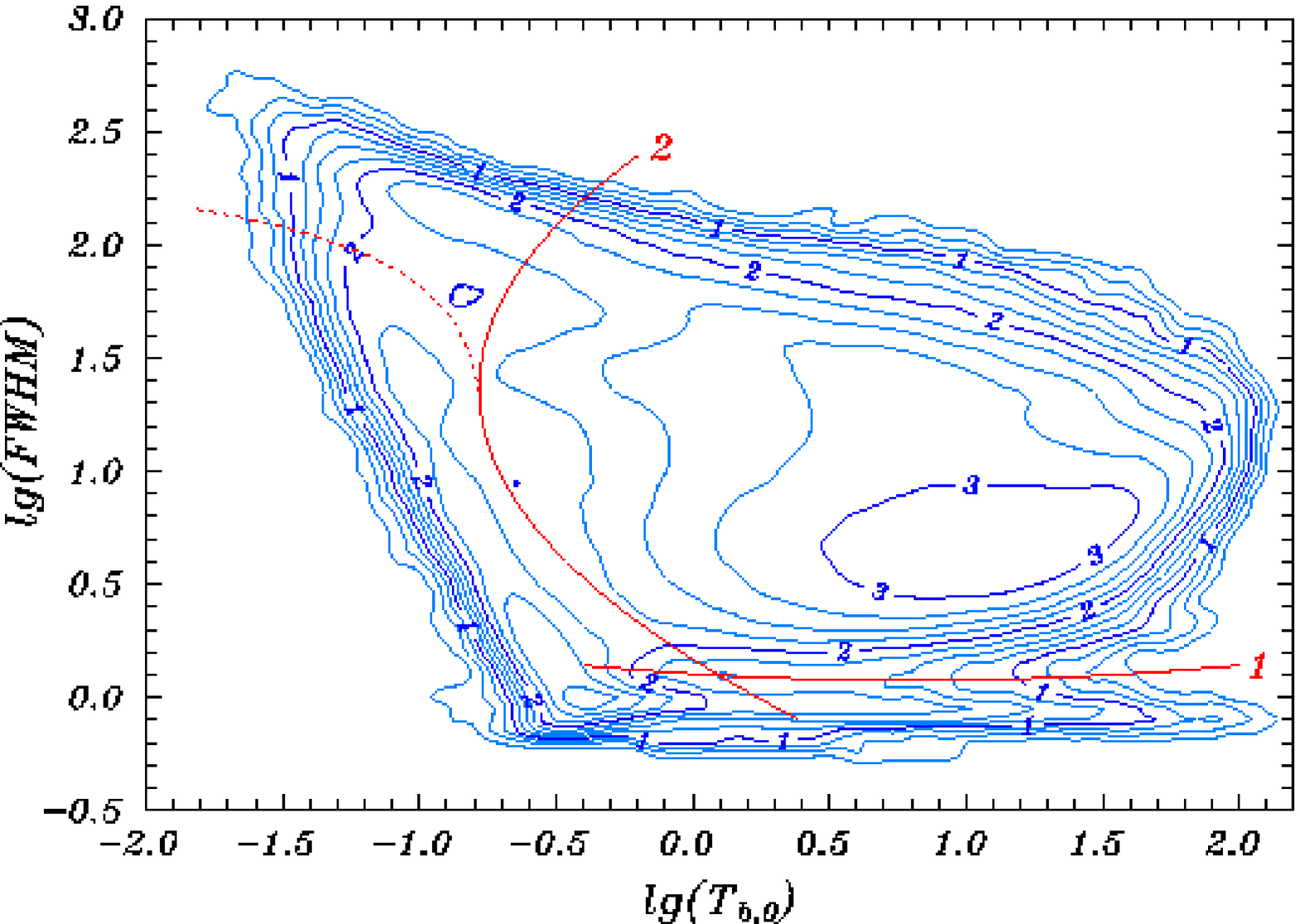,width=73mm,clip=}
         \vskip 1mm
         \captionb{8}{Frequency distribution of the parameters of all
             Gaussians corresponding to profiles at galactic latitudes
             $|b| < 30\degr$.}
         \vskip 5mm
         \psfig{figure=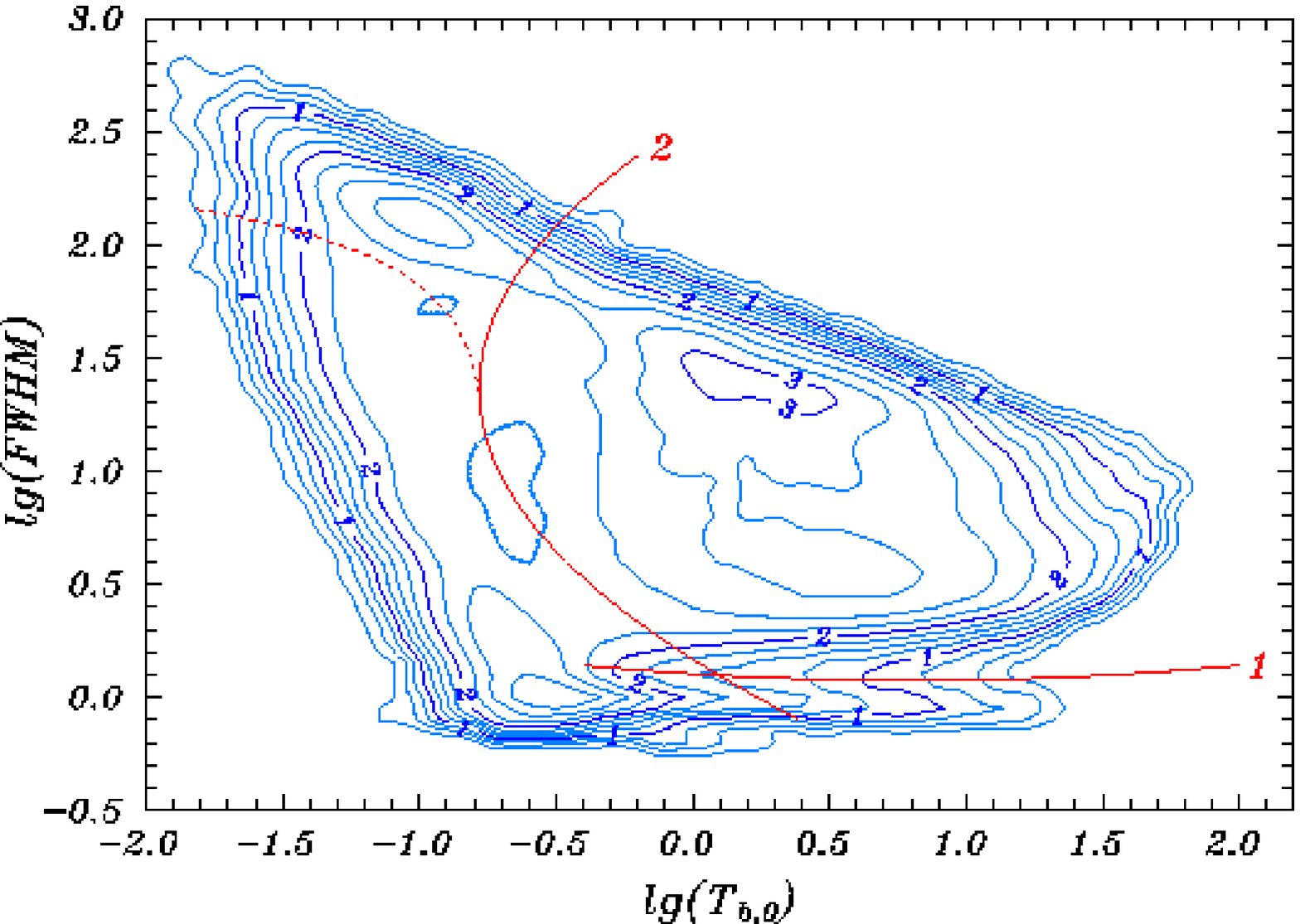,width=73mm,clip=}
         \vskip 1mm
         \captionb{9}{The same as Fig.~1, but for the published
            version of the LDS. The plot is scaled to the same number
            of profiles as in Fig.~1.}
      \end{wrapfigure}
      So far we have discussed mainly the data for the regions at $|b|
      \ge 30 \degr$, where the hydrogen profiles are simpler than near
      the galactic plane. At the same time, we have pointed out that
      the Gaussian analysis allows us to characterize general
      properties of the profiles from region to region in the sky and
      to draw conclusions based upon similarities and differences in
      profile shapes. Therefore, it is interesting to check how
      different the $(\lg(T_\mathrm{b0}),~\lg(\mathrm{FWHM}))$
      distribution of Gaussians at low galactic latitudes is in
      comparison to Fig.~1. The corresponding results are presented in
      Fig.~8. We can see that the main difference between Figs.~1 and 8
      is a greater extent of the distribution towards the stronger and
      wider Gaussians in Fig.~8. This means that at low galactic
      latitudes many Gaussians are higher and/or wider than at high
      latitudes~-- the property caused by a complex superposition of
      low optical density gas in and near the galactic plane, which
      could not allow us to distinguish in many cases the contribution
      of every single concentration of \ion{H}{i}.

      However, despite the presence of large numbers of Gaussians,
      representing the total emission of unknown numbers of actual gas
      concentrations with partly unknown properties, Figs.~1 and 8 are
      in general rather similar. At least both of these plots can be
      used to the same extent for the separation of the Gaussians,
      arising most likely from the real \ion{H}{i} emission from those
      probably caused by different observational and reductional
      problems and there is even no need to considerably change the
      selection criteria described above. In both figures we can also
      see similarities in the distribution of the Gaussians with
      realistic parameters: two main concentrations in the width
      interval $3 < \mathrm{FWHM} < 35~\mathrm{km\,s}^{-1}$. Therefore,
      it is possible to obtain from the near plane data at least some
      results similar to those obtainable at higher latitudes.

      When for the testing of the decomposition program we used the
      original profiles as described above, it is also interesting to
      check how much the decomposition results are affected by the
      averaging of the re-observed profiles and by re-gridding of the
      whole survey. Therefore, in Fig.~9 we present as an example an
      analog of Fig.~1 for the published version of the LDS. We can see
      that this version of the plot is rather similar to that of
      Fig.~1. The most significant difference is that in the
      decomposition of the LDS Atlas data the resulting Gaussians are
      somewhat weaker than in the case of original profiles. This is
      best visible if we compare the locations of the distribution
      maximums around $\lg (T_\mathrm{b0}) \approx 0.5$ and $\lg
      (\mathrm{FWHM}) \approx 0.6$ in Figs.~1 and 9 and this may be
      explained as a result of interpolation between profiles with
      slightly differing locations of the line peaks~-- these peaks
      have been smoothed down. The decomposition of the Atlas data
      contains also larger numbers of weak Gaussians and the weakest
      components are weaker than in the case of the original LDS.
      However, this is a rather natural result, as both, re-gridding
      (interpolation) and averaging of the observed profiles reduce the
      mean noise level of the results and force the decomposition
      program to fit more weak Gaussians to the data.

      The comparison of Figs.~1, 8 and 9 also illustrates the
      uncertainties of the selection criterion 2 for the widest
      Gaussians. When for widths $\lg (\mathrm{FWHM}) \la 1.35$ the
      line selected on the basis of Fig.~1 is more or less acceptable
      also for other cases, above this value for $|b| < 30\degr$ the
      dashed line seems to be more acceptable and for the Atlas data
      both versions of the line 2 are rather arbitrary.

   \sectionb{4}{LAB}

      \begin{wrapfigure}[18]{r}[0pt]{75mm}
         \vskip -4mm
         \psfig{figure=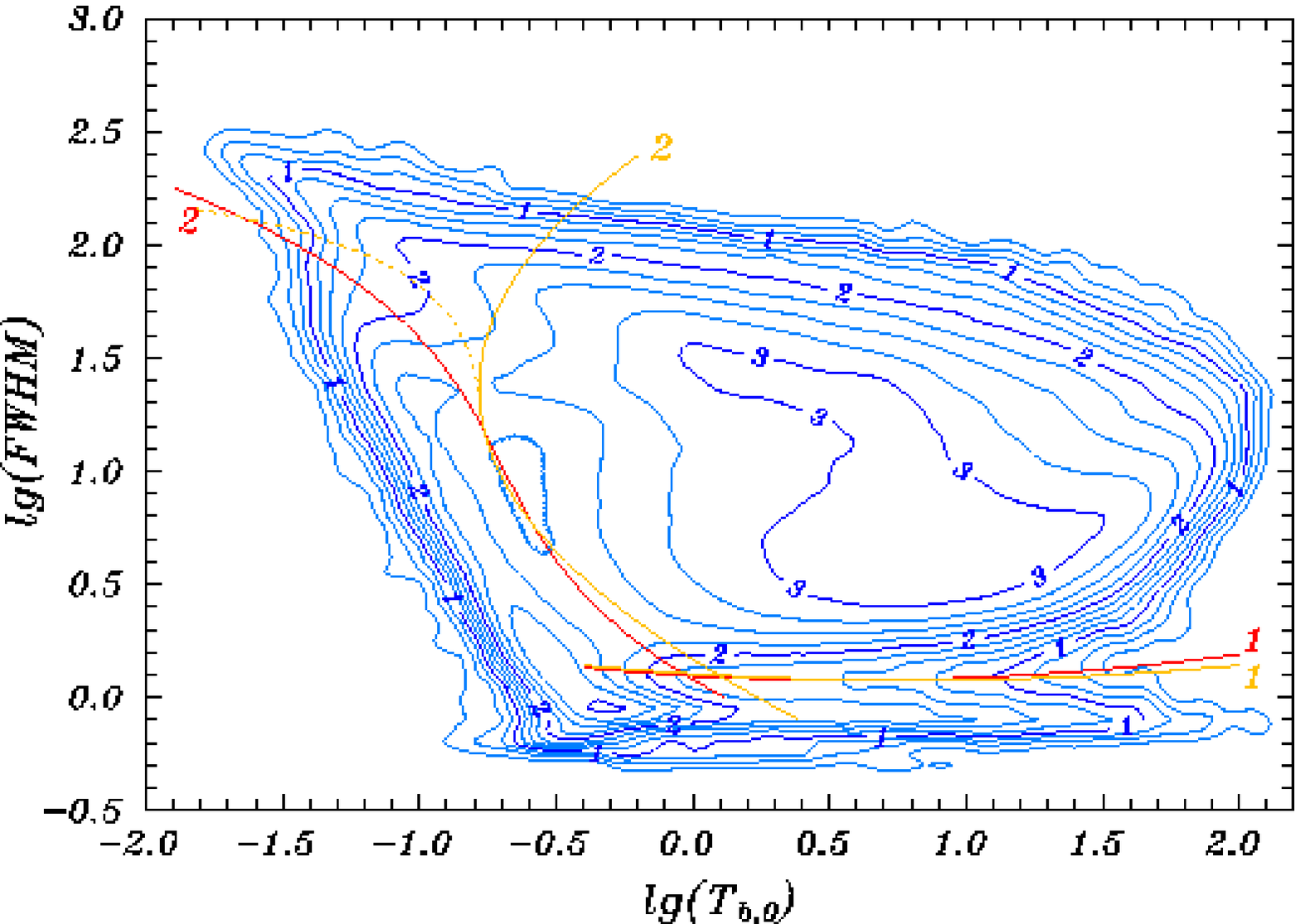,width=73mm,clip=}
         \vskip 1mm
         \captionb{10}{The same as Fig.~1, but for all profiles of the
            LDS2. The selection criteria derived for the LDS are still
            indicated by solid and dashed orange lines, but those used
            for the LAB are in solid red lines.}
      \end{wrapfigure}
      To cover the total sky, this compilation combines a revised
      version of the LDS with the IAR survey. We discuss the properties
      of both surveys individually.

   \subsectionb{4.1}{LDS2}

      When discussing the results for the original version of the LDS,
      we identified in the $(T_\mathrm{b0},~\mathrm{FWHM})$
      distribution three regions of the Gaussians, corresponding most
      likely to different problems, occurring during the observations
      and data reduction. These Gaussians represented the noise,
      radio-interferences, increased uncertainties near the profile
      edges and problems in baseline determination or in
      stray-radiation corrections. For the LDS2, included into the
      LAB, the stray-radiation corrections and the baselines were
      recalculated (Kalberla et al. 2005) and therefore it is
      interesting to compare the decomposition results for the LDS and
      the LDS2. The $(T_\mathrm{b0},~\mathrm{FWHM}$) distribution of
      the Gaussians obtained from the decomposition of the LDS2 is
      given in Fig.~10. In this case, all the profiles of the survey
      are used for the plot and therefore it corresponds to the sum of
      the distributions given in Figs.~1 and 8. From the comparison of
      these distributions we can see that in the LDS2 the widest
      Gaussians obtained for the LDS are missing, but there are no
      considerable changes in other regions of problematic Gaussians.
      This may be considered as an indication that the recalculation of
      the stray-radiation corrections and baselines has improved the
      results.

      \vbox{
         \vskip 3mm
         \centerline{\psfig{figure=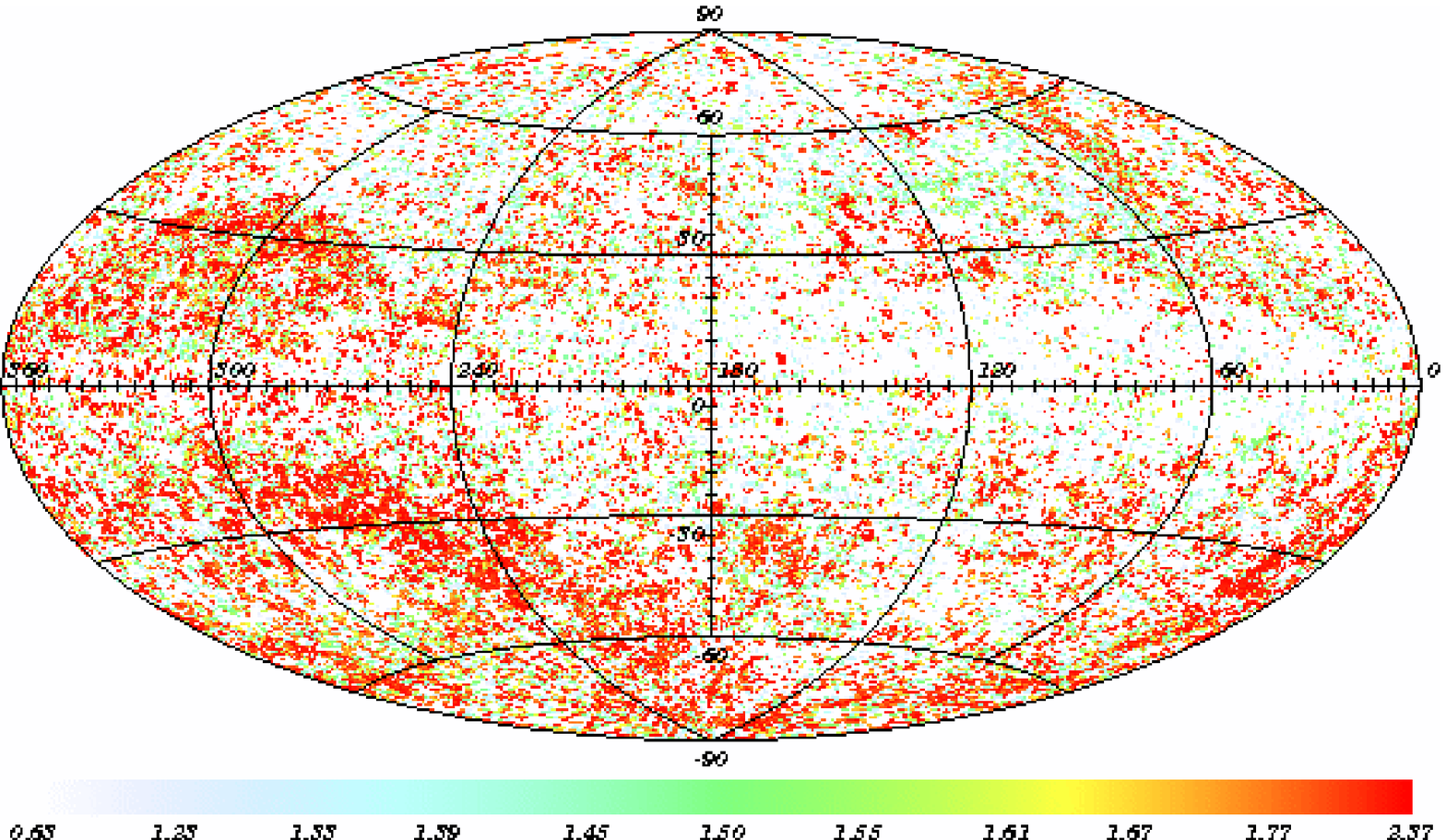,width=125mm,angle=0,clip=}}
         \vskip 1mm
         \captionb{11}{The sky distribution of the widest weak
            Gaussians of the LDS2 in galactic coordinates. The
            color-scale represents the width (in units of $\lg
            (\mathrm{FWHM})$) of the widest Gaussian obtained for a
            given sky position, and the gradation is chosen to enhance
            the contrast for the Southern sky.}
      }
      \vspace{3mm}
      To further check the situation with the wide Gaussians, we
      present in Fig.~11 for the LAB the distribution of the widest
      Gaussians in the sky. This picture is based on the same selection
      criteria as used for Fig.~6. Only, to stress some Southern sky
      features, the center of the color-scale is downshifted to
      $\lg(\mathrm{FWHM}) = 1.5$, but this has no considerable effect
      on the appearance of the Northern sky. When compared to Fig.~6,
      it is obvious that the checkered pattern, which was so
      conspicuous in the case of the LDS has nearly disappeared in the
      LDS2, and therefore we may conclude that the stray-radiation
      corrections and base-line determinations for the LDS2 have been
      made at least much more homogeneously than for the LDS.
      Concerning Fig.~10, this also means that in the case of the upper
      part of the selection criterion 2 the solid line preferred for
      the LDS seems now rather obsolete and the dashed curve is a much
      more attractive choice. Therefore, at least on the basis of the
      distribution of the Gaussian parameters most of the weak wide
      components seem to correspond to some real population of galactic
      \ion{H}{i}~-- most likely the HVDHG, discussed by Kalberla et al.
      (1998).

   \subsectionb{4.2}{IAR}

      When we started the decomposition of the IAR, the first results
      were rather disappointing and surprising: on an average, we got
      per every profile about 1.65 times more Gaussians than in the
      case of the LDS or the LDS2 and most of these Gaussians were
      relatively narrow and weak. This was exactly the behavior that
      may be expected from the decomposition program, if the estimates
      of the noise level of the profiles are too low. In this case the
      program tries to reduce the $\mathrm{rms}$ of the residuals below
      the actual noise level of the survey and the only way to achieve
      this is to add into the decomposition many weak narrow
      components, which represent the strongest peaks of the
      observational noise. Therefore, we first checked our
      determination of the noise level in signal-free regions of the
      profiles. We used several different algorithms for estimating
      this noise level (they have been described in detail in Paper I)
      and concluded that the results of different methods agreed very
      well with each other and with the estimate for the final mean
      $\mathrm{rms}$ noise of the database given by Bajaja et al.
      (2005). From this we concluded that the problem cannot lie in the
      determination of the noise level of the signal-free parts of the
      profiles.

      \begin{wrapfigure}[21]{r}[0pt]{75mm}
         \vskip -4mm
         \psfig{figure=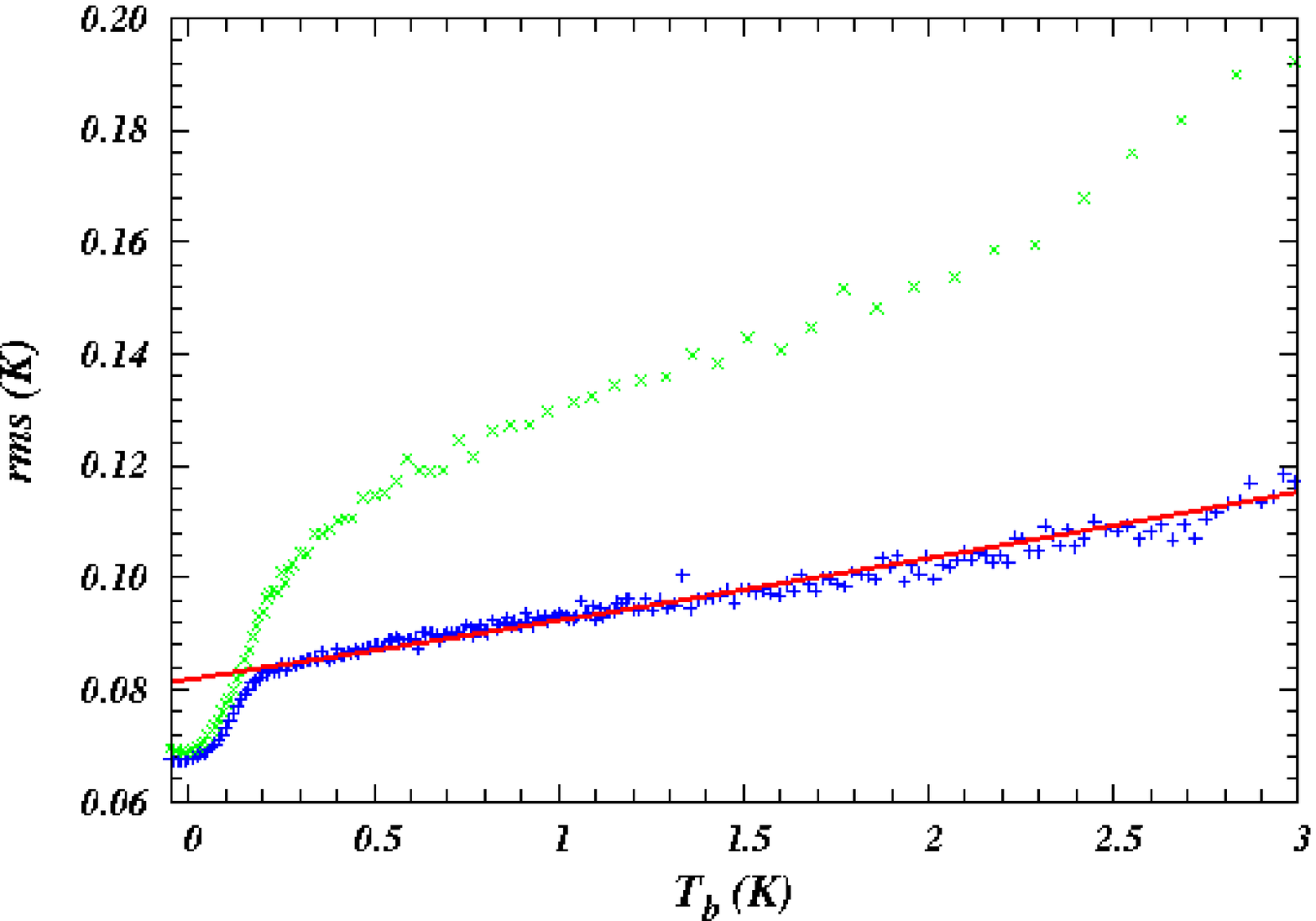,width=73mm,clip=}
         \vskip 1mm
         \captionb{12}{The dependence of the total uncertainties
            (green crosses) in channel values and of the observational
            noise (blue pluses) on the signal strength. The thick solid
            red line indicates extrapolation of the noise level in
            regions containing the line emission into the signal-free
            region.}
      \end{wrapfigure}
      Another possibility was that the problems may hide in the usage
      of the radiometer equation for the description of the noise
      strength dependence on signal intensity. To get the first
      insight, we compared channel by channel the re-observed profiles
      (the detailed description of this procedure is given in Section
      2.1. of Paper I). To reduce the role of uncertainties in the
      brightness temperature calibration and other possible scaling
      problems, we first compared the average channel values inside the
      usable velocity range of profiles at the same sky positions and
      rejected all results for the positions where the dispersion of
      these averages for different profiles was more than
      $0.05~\mathrm{K}$ (selecting even a smaller limit did not
      considerably change the results). Next we normalized all profiles
      at the same sky position to the same average channel value and
      compared all possible pairs of profiles at a given sky position
      channel by channel. In this comparison we used the average of
      corresponding channel values from different profiles as an
      indication of the signal level and their difference as an
      indication of the noise level. The results for small signal
      strengths are given with crosses in Fig.~12. We can see a rather
      unexpected behavior. Where the signal is missing, the results
      once again agree very well with the mean noise estimates for the
      survey. However, already for the $0.5~\mathrm{K}$ signal the
      uncertainties in channel values have increased nearly twice and
      only after this the uncertainties grow more or less linearly with
      the signal, as expected from the radiometer equation.

      What is described above, is a rather direct estimation of the
      uncertainties in channel values at different levels of the
      signal. However these estimates contain not only observational
      noise, but they may also have been considerably increased by
      uncertainties in data reduction procedures (baseline,
      stray-radiation etc.). To study the pure noise, we must reject
      other sources of differences between the reobserved profiles. To
      some extent, this can be done by smoothing every profile and
      taking the smoothed version of the profile as an estimate of the
      signal behavior and the differences between original and smoothed
      versions as an estimate of the noise. For such smoothing we used
      the Savitzky-Golay filters of the second degree with different
      window widths. In Fig.~12 the plus signs show the results for
      (2,2,2) filter. Once again, the results for signal-free regions
      agree well with other estimates, but in the region of signal
      strengths of about $0.02 - 0.5~\mathrm{K}$ there is a rapid
      increase in noise and only after this the noise behaves more or
      less as predicted by the radiometer equation. Of course, now the
      estimates of the noise strength in the regions containing a
      signal are considerably lower than those, obtained from the
      reobserved profiles, but these differences are in good agreement
      with the expectations by Bajaja et al. (2005) that ``the
      necessary interpolation of the baseline leads to uncertainties
      which are enhanced by a factor of 2 to 3 over the typical
      $\mathrm{rms}$ uncertainties of $0.07~\mathrm{K}$ as determined
      outside the regions with line emission''.

      \begin{wrapfigure}[19]{r}[0pt]{75mm}
         \psfig{figure=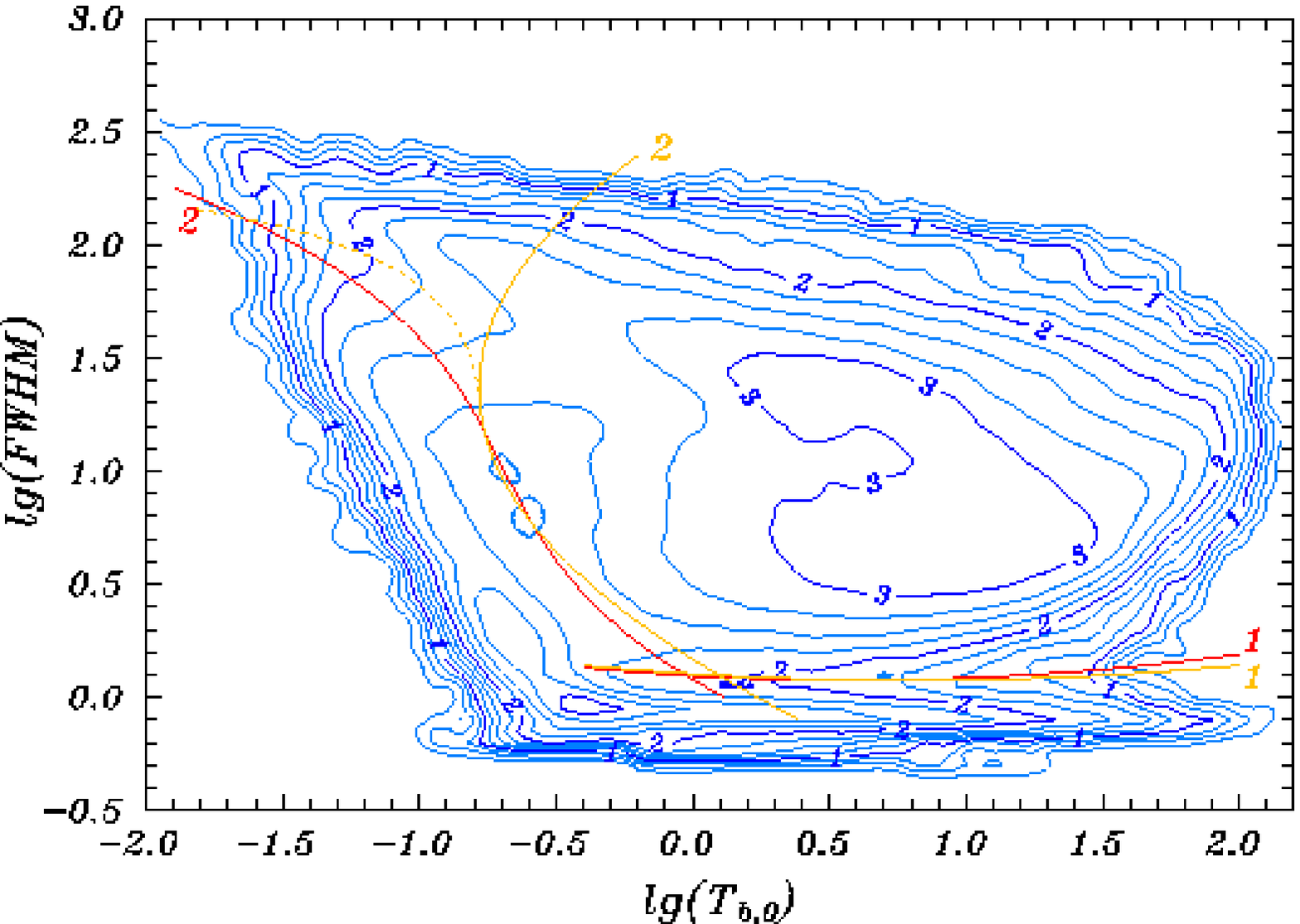,width=73mm,clip=}
         \vskip 1mm
         \captionb{13}{The same as Fig.~10, but for the IAR. The plot
            is scaled to the same number of profiles as in Fig.~10.}
      \end{wrapfigure}
      On the basis of the described results, we decided to slightly
      modify the noise estimates used in the decomposition program. We
      still obtain the main estimate of the noise strength from the
      signal-free regions of the profiles, but for regions with
      \ion{H}{i} emission we allow for a 16\% higher noise level. Such
      a model is indicated in Fig.~12 by the solid line. The 16\%
      increase is chosen as a conservative value from the results with
      different smoothing filters. This estimate is not very precise,
      but we believe that the actual value cannot be considerably
      smaller, but according to some results may be even somewhat
      larger (up to about 21\%). We also rechecked the LDS data for the
      presence of such a jump and found no need for introducing a
      similar correction in this case. After including the described
      16\% correction into the decomposition process of the IAR data,
      the results become much more similar to those of the LDS2. In the
      case of the IAR, there are still on an average 14\% more
      Gaussians per profile than for the LDS2, but this may be natural,
      as there is also about 27\% more \ion{H}{i} per profile in the
      Southern sky than in the Northern sky.

      \begin{wrapfigure}[18]{r}[0pt]{75mm}
         \psfig{figure=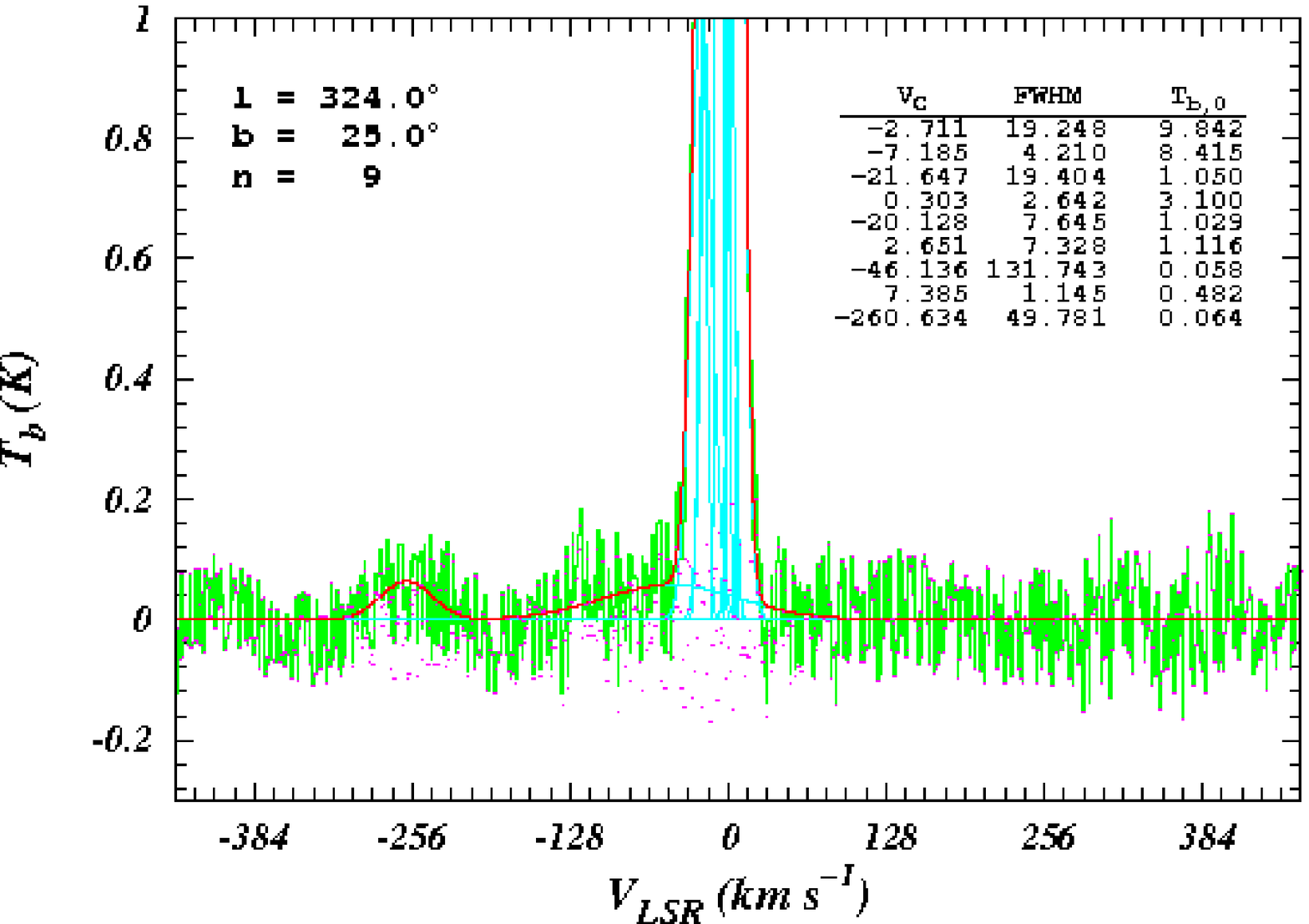,width=73mm,clip=}
         \vskip 1mm
         \captionb{14}{An example of the IAR profile with bad
            base-line. The notation is the same as in Fig.~2.}
      \end{wrapfigure}
      The $(T_\mathrm{b0},~\mathrm{FWHM}$) distribution of the
      Gaussians obtained from the decomposition of the IAR is given in
      Fig.~13. When comparing with Fig.~10, we can see that now the
      numbers of weak narrow Gaussians are in both cases rather
      similar, but for the IAR there are more strong narrow Gaussians
      than in the case of the LDS2. Most likely these are the
      interference induced profile components mentioned also by Bajaja
      et al. (2005). Due to the larger number of Gaussians per profile
      in the Southern sky the overall extent of the distribution in
      Fig.~13 is also wider than in Fig.~10, but the general shapes of
      the isolines are rather similar, except in the region $-1.0 \la
      \lg(T_\mathrm{b0}) \la -0.5$ and $1.5 \la \lg(\mathrm{FWHM}) \la
      2.0$, where the IAR seems to contain considerably more Gaussians
      than the LDS2. This is visible also from Fig.~10, where the
      Southern sky shows once again the quadrangular patterns discussed
      in connection with the baseline problems of the LDS and mentioned
      also by Bajaja et al. (2005). However, in this case the
      quadrangular pattern is visible in narrower Gaussians than in the
      LDS, indicating that in the IAR at least some baseline defects
      are smaller in their frequency extent (Fig.~14).

   \sectionb{5}{CONCLUSIONS}

      In this paper we have mainly used the high galactic latitude part
      $|b| \ge 30\degr$ of the Leiden/Dwingeloo Survey of galactic
      neutral hydrogen by Hartmann \& Burton (1997) to demonstrate how
      the Gaussian analysis could be used for statistical cleaning of
      the observed \ion{H}{i} profiles from observational noise and
      different observational and reductional artifacts. The removal of
      most of the observational noise is achieved by the process of the
      Gaussian decomposition. The program searches in profiles the
      regions where the measured brightness temperatures are above the
      estimated noise level, and fits this excess with the Gaussian
      function of three free parameters. The Gaussians are added until
      the $\mathrm{rms}$ of the residuals becomes close to the initial
      noise level estimate of the signal-free regions of the profile
      (the detailed description of the decomposition process is given
      by Haud 2000). At this point the residuals are considered as pure
      observational noise and discarded from further consideration.

      We have demonstrated that not all Gaussians obtained in this way
      could be considered as representing the real \ion{H}{i} emission
      of the Milky Way. A considerable part of the obtained Gaussians
      are still due to different observational, reductional and
      decompositional problems, which have occurred during this
      process. However, we have demonstrated that by analyzing the
      distribution of the parameters of the obtained Gaussians, it is
      possible to further clean the results by distinguishing the
      components, modeling different artifacts from those representing
      the actual emission of the galactic hydrogen. Such a separation
      is easy for very narrow Gaussians, which represent the
      radio-interferences not found and removed during the reduction of
      the observed profiles, and the strongest random noise peaks
      misinterpreted by the decomposition program as possible signal
      peaks. Disregarding such Gaussians seems to be possible on the
      basis of their location in the line-width~-- intensity
      distribution alone.

      A considerable amount of somewhat wider weak Gaussians seem to be
      caused by the increased uncertainties in bandpass removal near
      the profile edges. In the direction of wider Gaussians this
      region smoothly transforms to the one dominated by Gaussians,
      most likely representing problems in the determination of the
      baselines of the profiles. As demonstrated by Fig.~3, at these
      widths our criterion based on line-widths and heights alone,
      becomes somewhat unreliable. There are considerable numbers of
      components with heights larger than allowed by our selection
      criterion, but still most likely not representing the real
      \ion{H}{i} emission. However, these Gaussians could easily be
      recognized from their velocities. These are the Gaussians with
      central velocities lying outside the velocity limits of the
      profile, actually used during the Gaussian decomposition. They
      are the components, modeling the cases of rising profile edges
      and their intensities are estimated from the small number of
      profile channels covering only the minor part of the region,
      where the corresponding Gaussian has considerable intensities.

      In the LDS the region of the widest weak Gaussians is most likely
      dominated by the baseline problems, but when dealing with these
      components their velocity information must also be considered as
      in the regions $-390 < V_C < -230~\mathrm{km\,s}^{-1}$ and $180 <
      V_C < 280~\mathrm{km\,s}^{-1}$ the selection criterion based on
      the line-widths and heights alone will discard also some
      information about HVCs. At the level of the present discussion it
      remains unclear, if and how Gaussians represent the high velocity
      dispersion halo gas reported by Kalberla et al. (1998). On the
      basis of the data presented in their paper, it seems that our
      selection criteria for the LDS discard most of this gas as well.
      At the same time, it seems to be impossible to decide on the
      basis of the single profile data, wether it contains the emission
      from HVDHG or a baseline problem.

      It is well known that near the galactic plane the \ion{H}{i}
      21-cm emission line profiles are so complicated that it is
      impossible to derive from the Gaussian analysis reliable
      conclusions on physical properties of the gas concentrations.
      Nevertheless, the comparison of Figs.~1 and 8 demonstrates that
      at least the selection criteria for discarding different
      artifacts are applicable in both cases without modifications.
      Moreover, despite clear differences in the distributions of the
      parameter values of the Gaussians, representing the \ion{H}{i}
      emission for regions $|b| \ge 30\degr$ and $|b| < 30\degr$, there
      are still some similarities, indicating that at least for some
      profiles in the region $|b| < 30\degr$ not all of the useful
      information may be completely "washed out" by velocity crowding,
      blending and other effects.

      The approach used for the LDS is applied also to the new LAB,
      where we may conclude that the baseline estimates for the LDS2
      part of the LAB have been made considerably more uniform, if
      compared to the LDS (the conspicuous ``chess-board'' sky has
      disappeared), but for the IAR the baseline may still be somewhat
      questionable. Also, the problems caused by radio-interferences
      may be more severe in the case of the IAR than for the LDS2.
      Moreover, we point out the strange behavior of the observational
      noise in the IAR, where only the estimates for signal-free
      regions are in agreement with those, published by the authors of
      the survey. The behavior of the noise in the regions with line
      emission corresponds to the one expected on the basis of the
      radiometer equation only if we accept that the $\mathrm{rms}$ in
      these regions is about $15-21$\% higher than the estimate derived
      from the emission-free regions. Only after taking into account
      such behavior of the noise estimates, it is possible to obtain
      for the IAR the decomposition results similar to those of the
      LDS or the LDS2.

      When comparing the shapes of the distributions of Gaussian
      parameters for different version of the surveys, we may conclude
      that for most cases the selection criteria for separation of the
      components, most likely not related to the emission of galactic
      \ion{H}{i}, are rather universal. The only exception is the
      region of the widest Gaussians, which in the case of the LDS
      seems to be dominated by the components caused by the badly
      defined baseline, but the situation may be different for the
      LAB. Therefore, considering also the uncertainties in the
      determination of the selection criteria, those based on the LDS
      and given by Eqs. (\ref{Eq2}) and (\ref{Eq3}), are to some extent
      applicable also to the LAB. However, mostly due to the
      differences in the situation with the widest Gaussians, those
      indicated in Figs.~10 and 13 with solid red lines are preferred
      for the LAB. Corresponding functional expressions are:
      \begin{equation}
         \lg(\mathrm{FWHM}) = 0.057*\lg(T_\mathrm{b0})^2 -
                              0.067*\lg(T_\mathrm{b0}) +
                              0.094 \label{Eq4}
      \end{equation}
      for line 1 and
      \begin{equation}
         \begin{array}{ll}
         \lg(T_\mathrm{b0}) = & -0.370*\lg(\mathrm{FWHM})^3 +
                                 1.132*\lg(\mathrm{FWHM})^2 -\\
                              & 1.567*\lg(\mathrm{FWHM}) +
                                 0.117\\
         \end{array}
         \label{Eq5}
      \end{equation}
      for line 2 respectively. These criteria are selected on the basis
      of the full LAB data (the LDS2 and the IAR combined). Eq.
      (\ref{Eq4}) may be applied irrespective of the velocities of the
      Gaussians involved, but Eq. (\ref{Eq5}) seems to be useful mainly
      for the relatively slow components in the velocity interval of
      about $|V_C| \la 150~\mathrm{km\,s}^{-1}$. At higher velocities
      the parameters of the Gaussians, describing the HVCs, some
      external galaxies and the survey artifacts may be rather similar
      and it is hard to decide on the basis of these parameters alone,
      what may be the actual source of the corresponding Gaussian.  Eq.
      (\ref{Eq5}) becomes also important at the extreme velocities near
      the profile edges where the probability of spurious components
      increases. The Gaussians with the central velocities outside the
      velocity range, used for the decomposition, must also be excluded
      as due to the uncertainties introduced by the bandpass removal.

      Finally, we would like to stress that all the discussion in this
      paper is statistical in its nature and the selection criteria
      presented could not be taken as a final truth for every
      particular profile and Gaussian component. As we hope, these
      criteria permit us to detect and reject most of the problematic
      cases described above and in this way to reduce the number of the
      undesirable components in the database. However, there are
      certainly cases, not detected by these criteria and also cases
      where important astrophysical signatures may be removed.
      Therefore, in the first order these criteria are useful for
      statistical work on the Milky Way \ion{H}{i}. However, they can
      also be used as first guiding lines for labeling the problematic
      profiles in the surveys. For example, in the case of multiple
      observations at the same position of the LDS2, the criteria
      described above have been used to select the "best" profile which
      is expected to be the least problematic in the sense of the
      presence of spurious features discussed in this paper. However,
      the final decision on the nature of such features must be made on
      the basis of the inspection of the actual profiles and other
      astronomical observations.

      \vskip 5mm

      ACKNOWLEDGEMENTS. We would like to thank W. B. Burton for
      providing the preliminary data from the LDS for testing the
      decomposition program prior the publication of the survey. We are
      also grateful to him for serving as a referee of this paper and
      for valuable discussions. A considerable part of the work on
      creating the decomposition program was done during the stay of U.
      Haud at the Radioastronomical Institute of Bonn University. The
      hospitality of the staff members of the Institute is greatly
      appreciated. We are thankful to E. Saar and J. Pelt for valuable
      suggestions and discussions. The project was supported by the
      Estonian Science Foundation grant no. 6106.

      \vskip 5mm

   \References
      \refb
         Bajaja~E., Arnal~E.~M., Larrarte~J.~J., Morras~R.,
         P\"oppel~W.~G.~L., Kalberla~P.~M.~W. 2005, \aap, 440, 767
      \refb
         Baker~P.~L., Burton~W.~B. 1979, \aaps, 35, 129
      \refb
         Braun~R., Burton~W.~B. 2000, \aap 354, 853
      \refb
         Burton~W.~B. 1966, BAN 18, 247
      \refb
         Burton~W.~B. 1992, in The Galactic Interstellar Medium, ed.
         D.~Pfenniger \& P.~Bartholdi, Saas-Fee Advanced Course 21, 1
      \refb
         Cappa de Nicolau~C.~E., P\"oppel~W.~G.~L., 1986, \aap 164, 274
      \refb
         Crovisier~J. 1981, \aap, 94, 162
      \refb
         Dickey~J.~M., Lockman~F.~J. 1990, \araa, 28, 215
      \refb
         Field~G.~B., Goldsmith~D.~W., Habing~H.~J., 1969, \apjl, 155,
         L149
      \refb
         Hartmann~L. 1994, The Leiden/Dwingeloo Survey of Galactic
         Neutral Hydrogen, Ph. D.--Thesis, Leiden Univ
      \refb
         Hartmann~L., Burton~W.~B. 1997, Atlas of Galactic Neutral
         Hydrogen, Cambridge Univ. Press, 10+236~pp
      \refb
         Hartmann~L., Kalberla~P.~M.~W, Burton~W.~B., Mebold~U. 1996,
         \aaps, 119, 115
      \refb
         Haud~U. 2000, \aap, 364, 83
      \refb
         Kalberla~P.~M.~W., Westphalen~G., Mebold~U., Hartmann~D.,
         Burton~W.~B. 1998, \aap, 332, L61
      \refb
         Kalberla~P.~M.~W., Burton~W.~B., Hartmann~D., Arnal~E.~M.,
         Bajaja~E., Morras~R., P\"oppel~W.~G.~L. 2005, \aap, 440, 775
      \refb
         Kaper~H.~G., Smits~D.~W., Schwarz~U., Takakubo~K., van
         Woerden~H. 1966, BAN 18, 465
      \refb
         Kulkarni~S.~R., Fich~M. 1985, \apj, 289, 792
      \refb
         Kulkarni~S.~R., Heiles~C. 1987, in Interstellar Processes, ed.
         D.~Hollenbach \& H.~A.~Thronson, Jr. (Dordrecht: Reidel), 87
      \refb
         Mebold~U. 1972, \aap, 19, 13
      \refb
         Mebold~U., Winnberg~A., Kalberla~P.~M.~W., Goss~W.~M. 1982,
         \aap, 115, 223
      \refb
         P\"oppel~W.~G.~L., Marronetti~P., Benaglia~P. 1994, \aap, 287,
         601
      \refb
         Shane~W.~W. 1971, Observations of Neutral Hydrogen in an
         Interior Region of the Galaxy and the Structure and Kinematics
         of the Scutum Spiral Arm, Ph. D.--Thesis, Leiden Univ
      \refb
         Takakubo~K., van Woerden~H. 1966, BAN, 18, 488
      \refb
         Verschuur~G.~L., Knapp~G.~R. 1971, \aj, 76, 403
      \refb
         Verschuur~G.~L., Knapp~G.~R. 1972, \aj, 77, 717
      \refb
         Verschuur~G.~L., Schmelz~J.~T. 1989, \aj, 98, 267
      \refb
         Verschuur~G.~L., Magnani~L. 1994, \aj, 107, 287
      \refb
         Verschuur~G.~L., Peratt~A.~L. 1999, \aj, 118, 1252
      \refb
         Verschuur~G.~L. 2004, \aj, 127, 394
\end{document}